\documentclass[twocolumn, showpacs, preprintnumbers, amsmath, amssymb, superscriptaddress, prb]{revtex4}

\usepackage{graphicx}
\usepackage{dcolumn}
\usepackage{bm}
\usepackage{amssymb}
\usepackage{multirow}

\def\ket#1{| {#1} \rangle}

\begin{document}

\title{Strain-tuning of quantum dot optical transitions via laser-induced surface defects}

\author{Cristian Bonato}
\affiliation{Huygens Laboratory, Leiden University, P.O. Box 9504, 2300 RA Leiden, the Netherlands}

\author{Evert van Nieuwenburg}
\affiliation{Huygens Laboratory, Leiden University, P.O. Box 9504, 2300 RA Leiden, the Netherlands}

\author{Jan Gudat}
\affiliation{Huygens Laboratory, Leiden University, P.O. Box 9504, 2300 RA Leiden, the Netherlands}

\author{Susanna Thon}
\affiliation{University of California Santa Barbara, Santa Barbara, California 93106, USA}

\author{Hyochul Kim}
\affiliation{University of California Santa Barbara, Santa Barbara, California 93106, USA}

\author{Martin P. van Exter}
\affiliation{Huygens Laboratory, Leiden University, P.O. Box 9504, 2300 RA Leiden, the Netherlands}

\author{Dirk Bouwmeester}
\affiliation{Huygens Laboratory, Leiden University, P.O. Box 9504, 2300 RA Leiden, the Netherlands}
\affiliation{University of California Santa Barbara, Santa Barbara, California 93106, USA}

\email{bonato@physics.leidenuniv.nl}

\begin{abstract}
We discuss the fine-tuning of the optical properties of self-assembled quantum dots by the strain perturbation introduced by laser-induced surface defects. We show experimentally that the quantum dot transition red-shifts, independently of the actual position of the defect, and that such frequency shift is about a factor five larger than the corresponding shift of a micropillar cavity mode resonance. We present a simple model that accounts for these experimental findings.
\end{abstract}

\maketitle

Self-assembled quantum dots (QD) have attracted much interest as systems that show atom-like properties in the solid state and can be integrated in semiconductor heterostructures and devices. Numerous applications have been devised in different fields, like nanophotonics, optoelectronics and quantum information.\\
In quantum optics, single quantum dots are often embedded in microcavities to enhance the interaction with weak light fields, either in the weak or strong coupling regimes of cavity quantum electrodynamics. Efficient and reliable single photon sources \cite{lounisRPP05, scheelJMO09, straufNP07,  forchelIOP10, heindelAPL10} have been demonstrated with a quantum dot in a microcavity, thanks to the enhancement in the spontaneous emission due to the Purcell effect. Moreover, quantum information schemes employing cavity quantum electrodynamics with quantum dots coupled to semiconductor microcavities have been proposed and implemented.\cite{imamogluPRL99, reithmayerNature04, rakherPRL09, xuPRB09, bonatoPRL2010} Such system can provide a scalable platform for hybrid quantum information protocols, in which photonic qubits are used for long-distance transmission and matter qubits for local storage and processing.\cite{ciracPRL97, vanenkPRL97}\\
In the case of epitaxially-grown self-assembled quantum dots, \cite{warburtonCP02, michlerBook03} a thin semiconductor film is deposited by molecular beam epitaxy on a semiconductor substrate with a different lattice constant. Due to lattice mismatch, elastic strain energy builds up in the process, which is minimized by the formation of strained islands at the surface. These islands are transformed in quantum dots capping them with a larger band-gap material. Epitaxial growth gives a very strong three-dimensional carrier confinement, which results in atom-like quantization of the energy levels. It has, however, the drawback that the quantum dot position and size are poorly controlled. Techniques have been developed to control the position of quantum dots using a nanohole as a nucleation center and then fabricating a cavity around \cite {schneiderAPL08} or to position a nanocavity around a randomly located emitter.\cite{doussePRL08, thonAPL09} The optical properties of a given dot are determined by its composition, size and the local strain: therefore it is not possible to have a deterministic control on the frequency of the quantum dot optical transition. This is a crucial problem for cavity quantum electrodynamics, since the frequency of the emitter and the cavity mode resonance must be matched with high precision.\cite{badolatoScience05}
Moreover, several quantum information applications based on photon polarization \cite{xuPRB09, bonatoPRL2010} require the cavity mode to be polarization-degenerate. In general, fabrication imperfections and residual strain break the symmetry of the microcavity and the fundamental cavity mode is split in two linearly-polarized submodes. Techniques are therefore needed to fine-tune the optical properties of microcavities and match the resonance frequencies of the emitter and the cavity. \\
Permanent or reversible frequency shifting of the resonances of photonic crystal cavities can be achieved with different tools: wet chemical digital etching\cite{hennessyAPL05}, photodarkening of a thin chalcogenide glass layer \cite{faraonAPL08} or a photo-chromic thin film \cite{sridharanAPL10} deposited on top of the device, atomic force microscope nano-oxidation of the cavity surface \cite{hennessyAPL06}, infiltration of liquids \cite{ericksonOE06, intontiAPL06} or absorption of xenon \cite{mosorAPL05}. Such techniques, however, only work for photonic crystals, where the cavity is on the surface of the device and can be easily accessed. For micropillar cavities, on the other hand, the cavity region is buried under a multilayered mirror structure and it is not accessible for fabrication tuning processes.\\
Alternatively, the quantum dot transition has to be shifted onto resonance with a cavity mode. By embedding the dots in a diode structure and applying a voltage, the optical transition frequency can be shifted via the quantum confined Stark effect.\cite{fryPRL2000, kistnerOE08} The Stark shift can be finely tuned, but is limited to a range of hundreds of $\mu$eV:  it is therefore most effective in combination with some other coarse tuning techniques. Temperature-tuning, either of the whole sample \cite{kirazAPL01} or of a local spot \cite{faraonAPL07} has been shown as an effective way to get energy shifts on the order of $1-2$ meV. Temperature variation has been shown to have a stronger effect on the quantum dot transition (40 $\mu$eV/K, dominated by the temperature-dependence of the bandgap) than on the cavity mode (5 $\mu$eV/K, dominated by the temperature dependence of the refractive index) \cite{reithmayerNature04}. The available temperature range, however, is limited to $50$ K, beyond which the dot luminescence quenches and is affected by phonon-induced dephasing.\\
Lasers can also be used to tune quantum dot optical transitions \cite{rastelliAPL07} by altering their composition through local annealing: the heat caused by a laser beam creates temperatures sufficiently high that the the indium atoms contained in the QDs start to intermix with the gallium atoms in the surrounding matrix, resulting in a blue-shift of the quantum dot transition.\\
Strain also affects the optical properties of materials and can be exploited for tuning purposes \cite{obermullerAPL99, zanderOE09, bryantPRL10, seidlAPL06} both of quantum dot transitions and cavity mode resonances. Recently, it was shown that the strain created by laser-induced surface defects can be used to fine-tune the optical properties of semiconductor microcavities \cite{doornAPL96a, doornAPL96b, bonatoAPL2009}. Focusing a strong laser beam on a small spot, far away from the cavity center so that the optical quality of the device is not degraded, the refractive index can be changed so that the fundamental cavity mode can be made polarization-degenerate and tuned to a different absolute wavelength. Moreover, we showed \cite{gudatAPL11} that a careful combination of isotropic and anisotropic strain can be used to tune a quantum dot optical transition into resonance with a polarization-degenerate cavity mode, without affecting the cavity mode degeneracy.\\

Here we provide experimental and theoretical support for the technique, focusing on the physics involved with strain-tuning of quantum dot transitions. In particular, in Section I we will show experimentally that:
\begin{enumerate}
  \item The quantum dot line always red-shifts, independent of the actual position of the hole with respect to the cavity region
  \item The frequency shift for the quantum dot optical transition is larger than the corresponding absolute shift of the cavity mode resonance
\end{enumerate}
In Section II we will discuss a simple two-band model which can explain these results, giving sufficient qualitative information on the involved physics.

\section{Experimental results}

We investigated a sample with quantum dots embedded in micropillar cavities, grown by molecular-beam epitaxy on a GaAs [100] substrate. The microcavity consists of two distributed Bragg reflector (DBR) mirrors, made by alternating $\lambda/4$ layers of GaAs and Al$_{0.9}$Ga$_{0.1}$As. Within the mirrors there is a $\lambda$-thick GaAs layer, embedding InGaAs/GaAs self-assembled quantum dots, below an AlAs layer. Trenches are etched down to the bottom DBR, leaving a circular pillar with a diameter of the order of $30 \quad \mu$m and the sample is placed in a steam oven to create an AlO$_x$ oxidation front in the AlAs layer, leaving a small un-oxidized area in the center, with a diameter of $3-5 \quad \mu$m.\cite{gudatNewAPL} The difference in the effective refractive index between the oxidized and un-oxidized regions provides gentle transverse optical confinement, which results in high-quality factors ($Q \sim 30000$) and small mode volumes \cite{gudatAPL11, stoltzAPL05}.
\begin{figure}[htbp]
\centering
\includegraphics[width=8 cm] {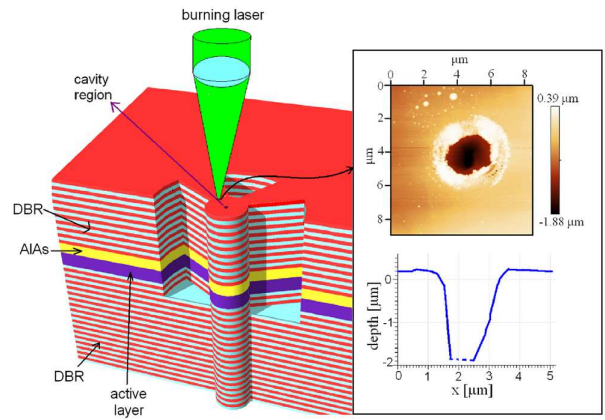}
\caption{Sketch of the micropillar samples with embedded quantum dots (see text for details). A laser beam (about 500 mW power, $\lambda = 532$ nm) is tightly focussed on the sample to create small defects, far away from the cavity area. Such defects (see AFM image in the inset) appear as holes, about $2$ $\mu$m wide and $2$ $\mu$m deep, with some material removed and deposited on the edges.}
\label{fig:AFMhole}
\end{figure}
By using micropillars defined by trench shapes, intra-cavity electrical gating of multiple devices is possible by the fabrication of a PIN-diode structure. An electric field can be applied through the PIN diode structure which enables controlled loading of electrons into the QDs \cite{warburtonNat00} and tuning of the emission wavelength of the QDs by quantum-confined Stark effect \cite{fryPRL2000}.

Defects can be created on the sample surface by a laser beam (about 500 mW power on a few $\mu$m$^2$ spot, $\lambda = 532$ nm) tightly focused on the structure for about 30 seconds by a high-NA aspheric lens $L_1$ (focal length $f_0 = 4.2$ mm, $NA = 0.6$). The whole process is done under vacuum in a helium-flow cryostat, at a temperature of $4K$. As shown in the AFM image in Fig. \ref{fig:AFMhole}, the material is locally melted and evaporated, leaving a hole which is approximately $2$ $\mu$m wide and $2$ $\mu$m deep. The hole is precisely positioned, with a $\mu m$ accuracy, onto the sample by means of an optical system consisting of the focusing lens $L_1$, mounted on a piezoelectric translational stage, and a second lens $L_2$ (focal length $f = 150$ mm) which images the sample onto a CCD camera. All holes are burnt at least $20 \mu m$ away from the un-oxidized cavity are, so no appreciable in the quality factor of the micropillars was detected.

\begin{figure}[htbp]
\centering
\includegraphics[width=7 cm] {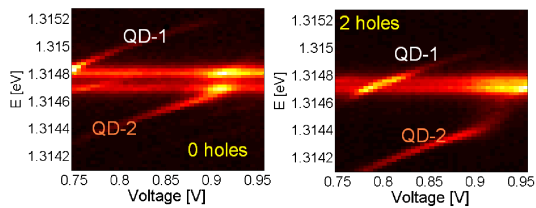}
\caption{Spectrally-resolved photoluminescence as a function of the applied bias voltage for increasing number of holes burnt around $20 \mu$m away from the dot. The energy-shifting lines correspond to quantum dots (Stark-shift), while the constant ones refer to the emission from cavity modes. Due to the effect of the holes burnt, the QD transitions (labeled as QD-1 and QD-2 in the figure) shift to lower and lower energies (about about $50$-$100$ $\mu$eV for each hole burnt).}
\label{fig:vScan}
\end{figure}

\begin{figure}[h]
\centering
\includegraphics[width=6cm] {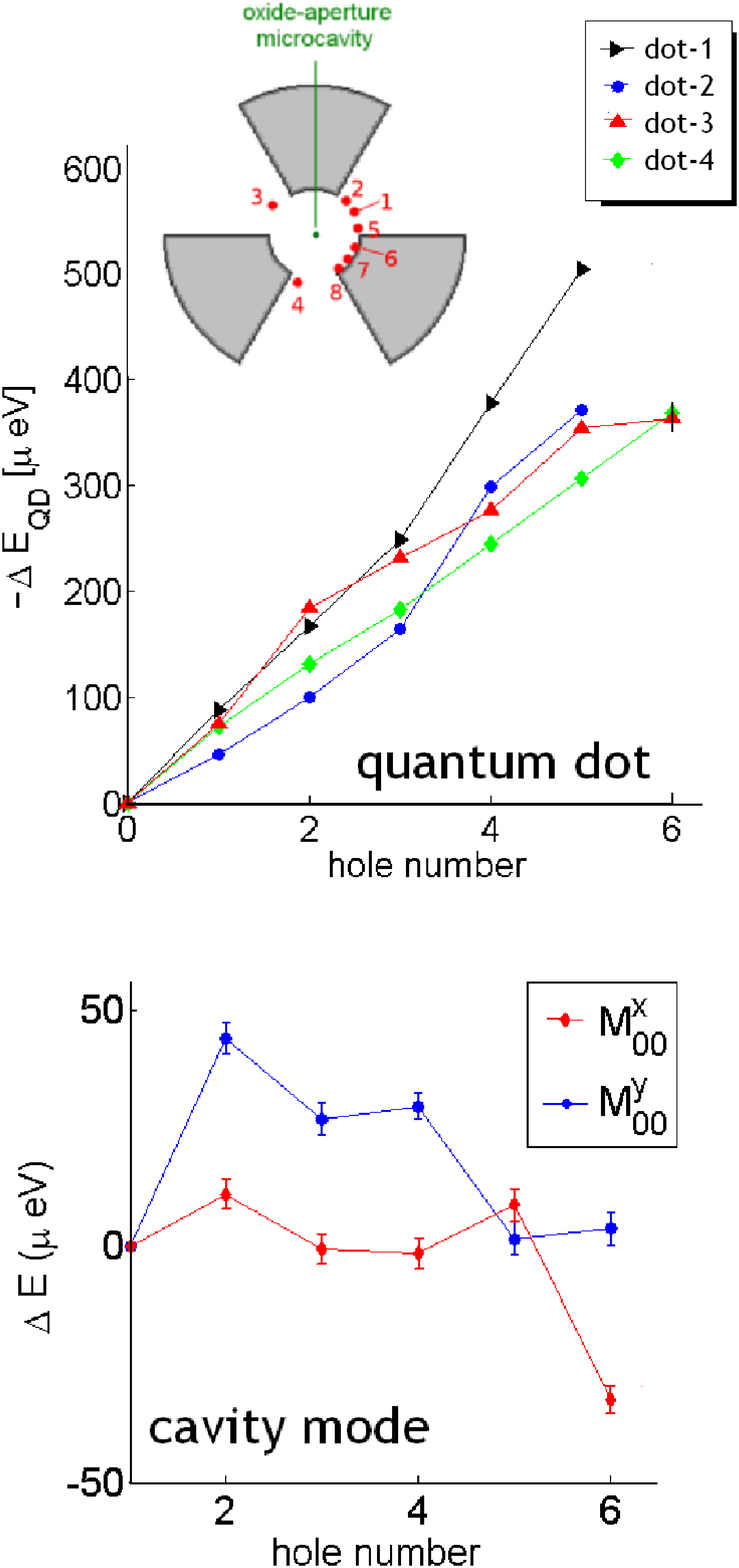}
\caption{The top plot shows the energy of four QD optical transitions with respect to their original energy for 6 holes burnt in total on the sample in different positions at the same distance from the cavity region (see inset for hole location). All transitions always red-shift ($\Delta E_{QD}$ keeps decreasing for each hole burnt), independently of the orientation of the hole position with respect to the cavity and the crystal axes. On the bottom plot the corresponding energy difference for the two orthogonally-polarized submodes M$_{00}^x$ and M$_{00}^y$ of the fundamental cavity mode with respect to their original (0 holes) resonance energy. M$_{00}^x$ and M$_{00}^y$ can either blue-shift or red-shift depending on the position of the hole. The dependence of the cavity mode shift on the hole position is described by the model in Section IIB.\\
Resonant frequencies were determined by a Lorentzian fit of the photoluminescence spectrum peaks, with an accuracy of the order of 4-5 $\mu$eV}
\label{fig:dotShift}
\end{figure}

After a hole has been burnt on the structure, the quantum dot optical properties are investigated by pumping the cavity region above the GaAs band-gap ($\lambda_p = 780$ nm, few $\mu W$ power on a few $\mu$m$^2$ spot size) and characterizing spectrally the photoluminescence. Tuning the voltage applied to the pin-diode, we switch between different charged states for the quantum dot while the corresponding optical transitions are frequency shifted by the Stark effect. Typical results are shown in Fig. \ref{fig:vScan}. We see that, by burning more holes on the sample, the QD transitions red-shift more and more, while the cavity modes are much less perturbed. Each hole shifts the dot transition of about $60$-$100$ $\mu$eV.  Fig. \ref{fig:dotShift} presents the energies of four QD transitions for holes burnt in different positions around the cavity area. The dot emission almost always red-shifts, independently of the angle $\theta$ along which the hole is burnt. In total, we only saw two cases of weak blue-shift out of tens of holes burnt. On the bottom graph in Fig. \ref{fig:dotShift} we plot the resonance energy for the two linearly-polarized submodes M$_{00}^x$ and M$_{00}^y$ of the fundamental cavity mode with respect to their original values. The frequencies of the two modes red-shift or blue-shift depending on the position of the hole.

\begin{figure}[h]
\centering
\includegraphics[width=7 cm] {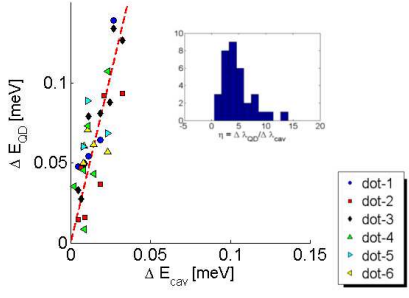}
\caption{Energy shifts for quantum dot and optical cavity mode for six different quantum dots. The dotted red line is a linear fit. On the inset histogram of the ratios between the shift of the dot and the shift of the cavity mode. On average the dot frequency shifts about $5$ times more than the cavity mode resonance.}
\label{fig:scatterPlot}
\end{figure}

In Fig.\ref{fig:scatterPlot} we plot the shift in the dot energy $\Delta E_{QD}$ as compared to the shift $\Delta E_{cav}$ of the cavity mode for each hole burnt for six different dots in the same sample. Since the fundamental cavity mode consists of two orthogonally-polarized submodes which are energy-shifted by $\Delta E_1$ and $\Delta E_2$, we take as the total shift $\Delta E_{cav} = |\Delta E_1 + \Delta E_2|$ as the absolute shift of the cavity mode. From the plot, the shift of the dot and the cavity mode appear to be correlated, and $\Delta E_{QD}$ is larger than $\Delta E_{cav}$. The average ratio $\eta$ between the shift of the dot and the cavity mode, calculated as the slope of the linear fit in Fig.\ref{fig:scatterPlot} is $\bar{\eta} = 4.5 \pm 0.3$. The error on $\eta$ is the error for the slope of the linear fit and the correlation coefficient is $r = 0.88$. Values for $\eta$ for different quantum dots are all compatible with each other within errors.

Summarizing, experimental data suggest that, due to hole-burning, the quantum dot optical transition always red-shifts, independently of the orientation of the laser-induced defect with respect to the crystal axes, and that this shift is about a factor $5$ larger than the corresponding shift for the cavity mode. In the following Section we will discuss a model to explain these observations.

\section {Theoretical model}
\subsection {Strain introduced by laser-induced surface defects}
Strain-tuning via the creation of small surface defects by means of a strong focused laser beam was first investigated as a tool to control the polarization properties of planar vertical-cavity semiconductor lasers by van Doorn et al. \cite{doornAPL96a, doornAPL96b}. They propose a theoretical model \cite{doornIEEE98} to interpret their  experimental data, showing that local heating in the vicinity of the device causes thermal expansion that results in the application of a controllable amount of strain.
Calculating the thermal expansion due to a point heat source in a bulk material, using the relations between the temperature of the point heat source and the elastic properties, one finds that, for a hole burnt at position $(x_0, y_0, 0)$, the $\sigma_{ij}$ stress component at position $(x, y, z)$ is given by ($x_i = x,y,z$) \cite{doornIEEE98}:
\begin {equation}
\sigma_{ij} = \gamma \frac{A_0}{r} \left[  \delta_{ij}-\frac{(x_i-x_i^{(0)}) (x_j-x_j^{(0)})}{r^2}\right]
\label {Eq:stress}
\end{equation}
where $\gamma = (C_{11}-C_{12})(C_{11}+2C_{12})/C_{11}$ and $x - x_0 = r \cos\theta \cos\varphi$, $y - y_0 = r \sin\theta \cos\varphi$, $z - z_0 = r \sin\varphi$.  $C_{ij}$ are the elastic constants of the material, while $A_0$ is a phenomenological coefficient, of dimension length, which depends on the laser power and on the thermal expansion coefficient and thermal conductivity of the material. $A_0$ is assumed to be positive in the case of compressive stress and negative for tensile stress. The $1/r$ factor gives a stronger effect near the position of the hole.\\
Above a certain laser power and burning time threshold, the strain effect becomes permanent and irreversible \cite{doornAPL96a}. In a previous experiment \cite {bonatoAPL2009}, we successfully explained the experimental data on the tuning of micropillar cavity modes by permanent laser-induced defects, assuming that the spatial distribution of stress is the same for reversible and permanent defects. Here we again assume that the stress given by permanent laser-induced defects has the form in Eq. \ref{Eq:stress}.

The applied stress induces strain (tensor $\varepsilon_{kl}$) in the crystal lattice, via the elastic compliance tensor, with components ${s_{ijkl}}$:
\begin{equation}
 \varepsilon_{ij} = s_{ijkl} \sigma_{kl}
\end{equation}
In matrix form, for a $4\bar{3}m$ trigonal crystal (like GaAs):
\begin{equation}
S = \left[
 \begin{array}{llllll}
S_{11} & S_{12} & S_{12} & 0 & 0 & 0\\
S_{12} & S_{11} & S_{12} & 0 & 0 & 0\\
S_{12} & S_{12} & S_{11} & 0 & 0 & 0\\
0 & 0 & 0 & S_{44} & 0 & 0\\
0 & 0 & 0 & 0 & S_{44} & 0\\
0 & 0 & 0 & 0 & 0 & S_{44}
 \end{array}
 \right]
\end{equation}
The coefficients $S_{ij}$ relate to the elastic constants $C_{ij}$ as: $S_{11} = (C_{11}+C_{12})/[(C_{11}-C_{12})(C_{11}+2C_{12})]$, $S_{12} = -C_{12}/[(C_{11}-C_{12})(C_{11}+2C_{12})]$ and $S_{44} = 1/C_{44}$.

Experimentally, the holes are burnt far from the position of the dot and the cavity center, not to degrade the optical quality of the device. Therefore, the angle $\varphi$, which describes how deep the dot position is in the $z$ direction with respect to the hole, can be considered to be very small ($\cos \varphi \simeq 1$).We can then calculate the components of the strain tensor to be:
\begin {equation}
\begin{array}{lll}
  \varepsilon_{xx} &\simeq &\left[1-\eta_1 \cos^2 \theta \right] \left( A_0/r \right)\\
  \varepsilon_{yy} &\simeq & \left[1-\eta_1 \sin^2 \theta \right]  \left( A_0/r \right)\\
  \varepsilon_{zz} &\simeq & A_0/r \\
  \varepsilon_{xy} &\simeq &- \eta_2 \sin\theta \cos\theta  \left( A_0/r \right)\\
\end{array}
\label {Eq:strain}
\end {equation}
where $\eta_1 = \left( 1+ 2 C_{12}/C_{11} \right)$ and $\eta_2 = (C_{11}-C_{12})(C_{11}+2C_{12})/(C_{11}C_{44})$. The strain components $\varepsilon_{yz}$ and $\varepsilon_{xz}$ are very small, since they are proportional to $\sin \varphi$.

\begin{figure}[h]
\centering
\includegraphics[width=7 cm] {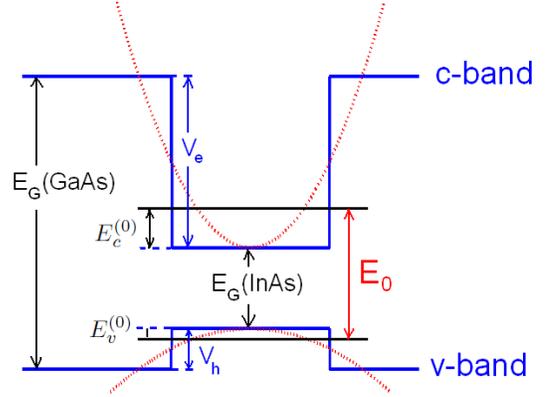}
\caption{Parabolic confinement band-edge approximation for the quantum dot potential.}
\label{fig:parabDot}
\end{figure}

\subsection {Effect on cavity modes}

The effect of laser-induced surface defects on the frequency of micropillar cavity modes has been thoroughly analyzed \cite{bonatoAPL2009}. Stress generated by the laser-induced defects (tensor $\sigma_{ij}$) creates strain in the semiconductor material (tensor $\varepsilon_{ij}$), via the elastic compliance tensor $S_{ijkl}$. Strain modifies the optical properties of the material through the elasto-optic tensor $p_{ijkl}$.  The change in the dielectric impermeability tensor $B_{ij}$ induced by a hole is  \cite{bonatoAPL2009}:
\begin {equation}
\delta B = \left[
\begin{array}{cc}
  \delta B_{xx} &   \delta B_{xy} \\
  \delta B_{xy} &   \delta B_{yy}
\end{array}
\right]
\end{equation}
where, for a $4\bar{3}m$ cubic crystal, $\delta B_{xx} = c_0 [\Pi_1 \sigma_{xx} - \Pi_2 \sigma_{yy}]$, $\delta B_{yy} = c_0 [- \Pi_2 \sigma_{xx} + \Pi_1 \sigma_{yy}]$, $\delta B_{xy} = \frac{p_{44}}{C_{44}} \sigma_{xy}$, $c_0^{-1} = (C_{11}-C_{12})(C_{11}+2C_{12})$, $\Pi_1 = p_{11} C_{11}-p_{12}C_{12}$ and $\Pi_2 = p_{11} C_{12}-p_{12}C_{11}$. Since $B_i = 1/n^2_i$, in the case of the small perturbation:
\begin {equation}
\frac{\Delta n_i}{n} \sim - n^2 \frac{\Delta B_i}{2}
\end {equation}
For a cavity with length $L$ and material refractive index $n$, the $m$-th resonant mode wavelength is $\lambda_m = 2 n L/m$, where the refractive indices $n_1$ and $n_2$ of the two submodes can be calculated from the eigenvalues $B_1$ and $B_2$ of the dielectric impermeability tensor $B$. A spatially uniform change $\Delta n$ in refractive index, results in a change in wavelength of the resonant mode $\Delta \lambda_m = \lambda_m (\Delta n /n)$. As an estimate of the isotropic shift of the cavity modes, we take the center of mass of the shift of the two resonance wavelengths $\Delta \overline{\lambda} = \Delta \lambda_1 + \Delta \lambda_2$. This quantity is proportional to the sum of the eigenvalues of $\delta B$, which corresponds to the trace of $\delta B$, and does not depend on its off-diagonal elements:
\begin{equation}
\begin {array} {lll}
\Delta \overline{\lambda} & \simeq  & \lambda_m \frac{n^2}{2} \left( \delta B_{xx} + \delta B_{yy} \right)\\
\qquad & \simeq & \lambda_m \frac{n^2}{2} \left( p_{11}+p_{12} \right) \left( 1-\frac{C_{12}}{C_{11}}\right) \left( \frac{A_0}{r} \right)
\end{array}
\end{equation}
The important point is that this quantity is independent of the angle $\theta$ along which the hole is positioned and is therefore a suitable definition to compare the effect of different holes on the cavity modes.
Using the values in Table \ref{Tab:material} for the elastic and elasto-optic coefficients  ($n \sim 3.5$ for GaAs), the total wavelength shift is:
\begin {equation}
\overline{\Delta \lambda} \mbox{[nm]} \sim  1100 \frac{A_0}{r}
\label{Eq:dLCavity}
\end{equation}
where $A_0$ is positive sign is for tensile strain (red-shift) and the negative for compressive strain (blue-shift). Measurements on the shift of the cavity mode suggest that the mode preferentially red-shifts when the cavity is far from polarization-degeneracy \cite{bonatoAPL2009}, which is an indication that tensile strain is being applied to the material by the laser-induced defects. Experimentally we have little control on the parameter $A_0$, so we cannot test this relation by its own. Instead, we need to compare it with the similar expression for the quantum dot shift.

\subsection {Effect on QD optical transitions}

In the $\mathbf{k}\cdot \mathbf{p}$ approximation for III-V semiconductors, we can treat the two-fold degenerate conduction bands separately from the valence bands, due to the large energy difference. The valence band consists of a doubly-degenerate band with angular momentum $j=1/2$ (spin-orbit split-off band) and two doubly degenerate bands with total angular momentum $j=3/2$. We neglect the spin-orbit split-off bands which, for typical semiconductors, are several hundreds meV separated from the four $j=3/2$ bands. Such bands are described by a $4$-by-$4$ Luttinger-Kohn hamiltonian \cite{chuangBook}. The states with larger band curvature ($m_j = \pm 1/2$) are called light-holes (LH), while the states with smaller band curvature ($m_j = \pm 3/2$) are called heavy-holes (HH). The Luttinger-Kohn hamiltonian, in the basis  $\lbrace \ket{m_z = \frac{3}{2}}, \ket{m_z =  \frac{1}{2}}, \ket{m_z =  -\frac{1}{2}}, \ket{m_z =  -\frac{3}{2}} \rbrace$ is:
\begin {equation}
\mathcal{H}_{LK} =
\left[
\begin{array}{cccc}
  P_k+Q_k & -S_k & R_k & 0 \\
  -S_k^* & P_k-Q_k & 0 & R_k \\
  R_k^* & 0 & P_k-Q_k & S_k \\
  0 & R_k^* & S_k^* & P_k+Q_k
\end{array}
\right]
\label {Eq:lk}
\end {equation}
where, for no strain:
\begin {equation}
\begin{array}{lll}
  P_k &= &\left(  \frac{\hbar^2}{2 m_0}\right) \gamma_1 \left( k_x^2 + k_y^2 + k_z^2  \right)\\
  Q_k &= &\left(  \frac{\hbar^2}{2 m_0}\right) \gamma_2 \left( k_x^2 + k_y^2 -2 k_z^2  \right)\\
  R_k &= &\left(  \frac{\hbar^2}{2 m_0}\right) \sqrt{3} \left[ -\gamma_2 \left( k_x^2 - k_y^2 \right) + 2i\gamma_3 k_x k_y  \right)\\
  S_k &= &\left(  \frac{\hbar^2}{2 m_0}\right) 2\sqrt{3} \gamma_3 \left( k_x - i k_y \right) k_z
\end{array}
\end {equation}
Here $\gamma_1$, $\gamma_2$ and $\gamma_3$ are the Luttinger parameters and $m_0$ is the free electron mass.

When strain is applied, the system is described by the Pikus-Bir Hamiltonian which has the same form of the Luttinger-Kohn Hamiltonian in Eq.\ref{Eq:lk}, but the coefficients are modified as $P = P_k + P_{\varepsilon}$, $Q = Q_k + Q_{\varepsilon}$, $R = R_k + R_{\varepsilon}$ and $S = S_k + S_{\varepsilon}$. Here:
\begin {equation}
\begin{array}{ll}
  P_{\varepsilon} &= -a_v \left( \varepsilon_{xx} +  \varepsilon_{yy}+  \varepsilon_{zz}\right)\\
  Q_{\varepsilon} &= (b/2) \left( \varepsilon_{xx} +  \varepsilon_{yy}- 2\varepsilon_{zz}\right)\\
  R_{\varepsilon} &= (\sqrt{3}b/2) \left( \varepsilon_{xx} -  \varepsilon_{yy}\right) - i d \varepsilon_{xy}\\
  S_{\varepsilon} &= d \left( \varepsilon_{xz} - i\varepsilon_{yz} \right)
\end{array}
\end {equation}
where $a_c$, $a_v$, $b$ and $d$ are the Pikus-Bir deformation potentials.

\begin{table}[h]
\caption {Material parameters used in the model \cite{chuangBook, watsonJLT04}}
\begin{center}
\begin{tabular}{lccc}
& & GaAs& InAs\\ \hline \hline
\multirow{3}{*}{elastic constants $[10^{10} N/m^2]$} & $C_{11}$ &$11.879$ & $8.329$ \\
& $C_{12}$ & $5.376$ & $4.526$\\
& $C_{44}$ & $5.94$& $3.964$ \\ \hline
\multirow{3}{*}{elasto-optic constants} & $p_{11}$ &-0.165 & -0.040\\
& $p_{12}$ & -0.140 & -0.035\\
& $p_{44}$ & -0.072& -0.010 \\ \hline
\multirow{3}{*}{Luttinger parameters} & $\gamma_1$ & 6.8 & 20.4 \\
& $\gamma_2$ & 1.9 & 8.3\\
& $\gamma_3$ & 2.73& 9.1 \\ \hline
\multirow{4}{*}{Pikus-Bir potentials [eV]} & $a_c$ & -7.17 & -5.08\\
& $a_v$ & 1.16& 1.00\\
& $b$ & -1.7& -1.8 \\
& $d$ & -4.55& -3.6\\
\hline \hline
\end{tabular}
\end{center}
\label{Tab:material}
\end{table}

In the following we will discuss a simplified two-valence-bands model, which accounts for the energy shift of the dot optical transition. Such a model is not expected to predict precisely the experimental results, but to give a phenomenological understanding of the physics behind the tuning process. More complex and accurate theoretical models \cite{andreaniPRB87, grundmannPRB95, liPRB96, tadicPRB02} have been developed to account for the effect of strain on the energy levels of semiconductor nanostructures. Although these models can be adapted to the problem we are considering, we believe that our simple model is accurate enough to understand qualitatively the results shown in the previous Section and to give sufficient guidance for the experimentalist when using laser-induced defects to strain-tune quantum dot samples.

\begin{figure}[h]
\centering
\includegraphics[width=7 cm] {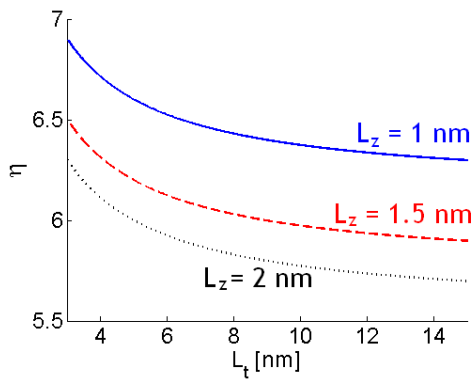}
\caption{Values of $\eta$, ratio between the frequency shifts of the quantum dot optical transitions and cavity mode resonance, for different dot dimensions.}
\label{fig:parabDot}
\end{figure}

\subsubsection {Two-valence-bands model}
In bulk III-V semiconductors, the heavy-hole and light-hole valence sub-bands are degenerate at $\mathbf{k}=\mathbf{0}$. In low-dimensional structures, however, the degeneracy is lifted by confinement and strain. The typical energy splitting for self-assembled quantum dots is on the order of a few tens of meV, which, in first-order approximation, allows neglecting the LH bands and just considering the conduction band and the HH valence band. In the spirit of trying to get a qualitative explanation of the experimental results, without complicated numerical simulations, we just consider the band-edge ($\mathbf{k}=\mathbf{0}$).\\
In the effective mass and envelope function approximations \cite{chuangBook}, we can consider the wave function for a single particle in a QD to be described by the product of a Bloch function $u_k (r)$, which has the periodicity of the atomic lattice, and an envelope function $f (r)$, which describes the amplitude modulation of the wave function that is imposed by the confinement potential: $\psi (r) = f(r) u_k (r)$.
The effective masses are $m^*_{e, t} = m^*_{e, z} =m_0/\gamma_e$ for electrons and $m^*_{h, t} = m_0/(\gamma_1+\gamma_2)$, $m^*_{h, z} = m_0/(\gamma_1-2\gamma_2)$ for heavy-holes, where $m_0$ is the free-electron mass.

We consider small dots, whose size is smaller than the corresponding bulk exciton radius ($\sim 35$ nm for InAs, $\sim 15$ nm for GaAs). In this case (strong-confinement approximation), electrons and holes can be considered as independent particles with energy primarily determined by the confinement potential, while the electron-hole Coulomb potential can be neglected or treated as a perturbation \cite {bryantPRB1988, franceschettiPRL1997, nazirPRB2005}.

Consider a quantum dot spherically symmetric in the $xy$-plane (width $2 L_t$) and with a depth $2 L_z$ along the growth axis $z$. Looking at the the band-edges, we get the potential well shown in Fig. \ref{fig:parabDot} for each direction.  At low temperature the energy gap is $E_g = 1.52$ eV for GaAs and $E_g = 0.42$ eV for InAs. We take the valence band offset to be $V_{h}^{(o)} = 0.25$ eV \cite{chuangBook}, which leaves $V_{c}^{(o)} = 0.87$ eV for the conduction-band offset.
In the strong confinement approximation, we can neglect the Coulomb interaction and take the energy of the exciton as:
\begin{equation}
E_0 = E_g^{(0)} \mbox{(InAs)} + E_v^{(o)} +E_c^{(o)}
\end{equation}
where $E_c^{(o)}$ and $E_v^{(o)}$ are the ground state energies for the potential wells for the conduction and valence potential wells.

To get a simple analytical solution we approximate the potential well with a three-dimensional parabolic potential \cite{nazirPRB2005}:
\begin {equation}
V_i (r) = \frac{1}{2} c_{i, t} \left( x^2 + y^2 \right)  + \frac{1}{2} c_{i,z} z^2 \quad i = e, h
\end{equation}
The coefficients $c_{i,j}$ can be found approximating the square finite well with the parabola:
\begin {equation}
\frac{1}{2} c_{i,j} L_{j}^2 = \frac{V_i}{2}
\end{equation}
which gives $c_{i, j} = V_i/L_j^2$ and $\Omega_{i, j} = (1/L_j)* \sqrt{V_i/m^*_{ij}}$. The index $i$ spans the conduction and valence bands ($i = e, h$), while $j$ identifies either the transverse coordinate in the $xy$-plane ($j = t$) or along the growth direction ($j = z$). $m^*_{ij}$ is the effective mass for the electron ($i=e$) or the heavy-hole ($i=h$) in the InAs potential well along the $j$ direction. The ground state energy for the parabolic potential well is $E_i^{(0)} = \frac{\hbar}{2} \left( \Omega_x + \Omega_y + \Omega_z \right) $, which gives:
\begin {equation}
\begin{array}{lll}
E_c^{(o)} &= &\frac{\hbar}{2}\sqrt{\frac{V_e}{m_0}} \left( \frac{2}{L_t}+\frac{1}{L_z} \right) \sqrt{\gamma_e}\\
E_v^{(o)} &= &\frac{\hbar}{2}\sqrt{\frac{V_h}{m_0}} \left[ \frac{2}{L_t} \sqrt{\gamma_1+\gamma_2} +\frac{1}{L_z}\sqrt{\gamma_1-2\gamma_2} \right]
\end{array}
\label{Eq:groundStates}
\end {equation}

The parabolic well approximation is excellent for the potential well in the $xy$-plane \cite{biolattiPRB02, rielAJP08}. Along the $z$ axis the potential well is more abrupt and the parabolic confinement approximation is not so good. This is hardly important, as we only look for a simple model to understand the physics and the order of magnitudes of the effects we experimentally observe.

Taking into account the strain induced by a laser defect, the band-edges are modified as $E_c = E_c^{(o)} + \Delta_e$ and $E_v = E_v^{(o)} + \Delta_h$, with:
\begin {equation}
\begin{array}{lll}
\Delta_e &= &a_c \varepsilon_H \\
\Delta_h &= &a_v \varepsilon_H + \frac{b}{2} \varepsilon_B
\end{array}
\end {equation}
Using the expression for strain derived in Eq. \ref{Eq:strain}, we get for the hydrostatic strain component $ \varepsilon_H = \varepsilon_{xx}+ \varepsilon_{yy}+ \varepsilon_{zz} = 2 (A_0/r) \left(1-C_{12}/C_{11}  \right)$ and for the biaxial component (assuming $\varphi \approx 0$) $ \varepsilon_B = \varepsilon_{xx}+ \varepsilon_{yy}-2\varepsilon_{zz} = - (A_0/r) \left(1+2 C_{12}/C_{11}  \right)$. Therefore:
\begin{equation}
\begin {array} {lll}
\Delta_e &= &2a_c \left(1-\frac{C_{12}}{C_{11}}  \right) \left( \frac{A_0}{r} \right) \\
\Delta_h &= & \left[ 2a_v \left(1-\frac{C_{12}}{C_{11}}  \right)-\frac{b}{2} \left(1+2\frac{C_{12}}{C_{11}}  \right) \right] \left( \frac{A_0}{r} \right)
\end{array}
\end{equation}

Note that the strain components that affect the conduction band and the heavy-hole valence band do not depend on relative angle $\theta$. This explains the data shown in Fig. \ref{fig:dotShift}, where the dot optical transition was shown to red-shift independently of the angle $\theta$ along which the hole was burnt.

The perturbation in the band-energies for the conduction and valence bands for GaAs and InAs has two consequences. First of all, the band-gap energy of InAs is modified as $E_g \mbox{(InAs)} = E_g^{(0)} \mbox{(InAs)} + \Delta E_G$, with:
\begin {equation}
 \Delta E_g = \Delta_e (\mbox{InAs}) -\Delta_h (\mbox{InAs}) = (a_c - a_v) \varepsilon_H - \frac{b}{2} \varepsilon_B
\end{equation}

Second, the depth of the potential wells is modified, due to the relative shift between GaAs and InAs band edges, giving a perturbation on the ground state energies of the potential wells (respectively $\Delta E_c$ and $\Delta E_v$). The confining potential for the electron is modified to $V_e = V_e^{(o)} + \delta_e *(A_0/r)$, with $\delta_e * (A_0/r) = \Delta_e (\mbox{GaAs}) - \Delta_e (\mbox{InAs})$, while the one for the holes is modified to $V_h = V_h^{(o)} + \delta_h *(A_0/r)$, with $\delta_h* (A_0/r) = \Delta_h (\mbox{GaAs}) - \Delta_h (\mbox{InAs})$.

Substituting the new expression for the confining potential in the ground-state energies in Eq. \ref{Eq:groundStates} and using a first-order Taylor expansion, since the perturbation due to a hole burnt is very small, we get:
\begin {equation}
\Delta E_i = \xi_i \delta_i \left( \frac{A_0}{r} \right) \quad i = c, v
\end {equation}
As shown above, the $\delta_i$ depend on the elastic properties of the materials, namely the elastic constants and the deformation potential. The $\xi_i$, on the other hand, depend on the band-structure:
\begin{equation}
\begin {array} {lll}
\xi_c &= &\frac{\hbar}{2} \left( \frac{1}{L_z} + \frac{2}{L_t} \right) \sqrt{\frac{\gamma_e}{m_o V_e^{(o)}}} \\
\xi_v &= &\frac{\hbar}{2} \left( \frac{1}{L_z} \sqrt{\gamma_1-2\gamma_2} + \frac{2}{L_t}\sqrt{\gamma_1+\gamma_2} \right) \sqrt{\frac{1}{m_o V_h^{(o)}}}
\end{array}
\end{equation}

Let's consider a quantum dot 3-nm thick and 12 nm wide ($L_z = 1.5$ nm and $L_t = 6$ nm). We get $\xi_e \sim 1.224$ and $\xi_h \sim 1.031$. Using the values in Table \ref{Tab:material}, we get $\Delta_e (\mbox{InAs}) \sim -4.638\mbox{eV} * (A_0/r)$ and $\Delta_h (\mbox{InAs}) \sim  2.791\mbox{eV} * (A_0/r)$. Therefore, in the case of compressive strain the band-gap energy for InAs increases as $E_g [\mbox{InAs}] = E_g^{(0)} [\mbox{InAs}] +7.43 \mbox{eV} * (A_0/r)$, while it decreases by the same amount for tensile strain.
The shift in GaAs conduction band is $\Delta_e (\mbox{GaAs}) \sim  -6.546 \mbox{eV} * (A_0/r)$, so that the change in the depth of the potential well is $\delta_e * (A_0/r) = -1.91 \mbox{eV} * (A_0/r)$. In the same way, the shift for the valence band is $\Delta_h (\mbox{GaAs}) \sim -2.678 \mbox{eV} * (A_0/r)$, so that the change in the depth of the corresponding potential well is $\delta_h * (A_0/r) =  0.113 \mbox{eV} * (A_0/r)$. For both the conduction and valence band, the depth of the potential well is increased by compressive strain and reduced by tensile strain. The effect is, however, much stronger for the conduction band than it is for the valence band.

Taking into account both the change in InAs energy gap and in the potential well depth, in the case of tensile strain, the considered quantum dot transition red-shifts due to hole-burning by an amount $\Delta E \sim -9.78 \mbox{eV} (A_0/r)$, which results in the following wavelength shift:
\begin {equation}
\Delta \lambda \mbox{[nm]} \sim 7800 \left( \frac{A_0}{r} \right)
\label{Eq:dLnm}
\end{equation}
To get an experimental value of about $100$ $\mu$eV for the typical dot transition shift given by one hole burnt, the value of $(A_0/r)$ is of the order of $10^{-5}$. Using this value the shift of the InAs bandgap for one hole burnt is around $75 \mu$eV. The change in the conduction-band confinement potential is around $\delta_e* (A_0/r) \sim -20\mu$eV and for the valence-band confinement well is $\delta_h* (A_0/r) \sim 1 \mu$eV.

Comparing with Eq. \ref{Eq:dLnm} with Eq. \ref{Eq:dLCavity}, we get:
\begin {equation}
\eta = \frac{\Delta \lambda_{QD}}{\Delta \lambda_{cav}} \sim 6
\end{equation}
In the case of compressive stress, $A_0$ is negative and the quantum dot transition blue-shifts by the same amount.

Experimental data, discussed in Section I, show that the quantum dot optical transition red-shifts. According to our model this suggests that the defect we burn in the sample induce tensile strain on the dots. We think that, by burning holes and removing material, we release some of the built-in compressive strain in the InAs quantum dot (due to the lattice-mismatched growth), which acts as an effective tensile strain. The relative magnitude $\eta$ of the effects for the quantum dot transition versus that for the cavity mode is plotted in Fig. 6 for different dot sizes. It clearly depends on $L_z$ and $L_t$ but the dependence is weak: for the values reported in the Figure, which cover emission energies ranging between 1 and 1.5 eV, $\eta$ is bounded between $5.5$ and $6.5$. Our analysis of the QD transition frequency shift can be considered then relatively robust with respect to variations in the dot size and size variations cannot be held responsible for the large spread in the experimental values for $\eta$ shown in Fig. 4. This claim is also supported by the fact the points corresponding to different dots are not aligned along lines with different slopes, which would be correlated to different dot sizes, but they are randomly spread.
The spread of the data points cannot be attribute either to a dependence on the angle $\theta$ along which the hole is burnt, since we find no correlation between dot shift and hole orientation (the correlation coefficient is only $r = 0.1$). This matches the theoretical predictions, since we defined the absolute cavity mode shift in Eq. 7 to be independent of $\theta$. Therefore, the spread of the data points should be attributed just to experimental errors and no physics seems to be related to it.
In conclusion, the predictions of our simple model are in good agreement with the experimental data shown in Fig. 3 and Fig. 4.

\subsubsection {Beyond the two-valence-bands model}
The model described in the previous Section can qualitatively explain the absolute red-shift of the transitions, but it is clearly inadequate for a full understanding of the effect of strain on the quantum dots.

First of all, the effect of valence-band mixing is neglected in a two-band model. In case the Luttinger-Kohn hamiltonian is diagonal ($R=S=0$) the eigenfunctions are either heavy-holes or light-holes states and the optical selection rules are such that a $\sigma_+$ photon creates a $\ket{\downarrow}$-electron and $\ket{\Uparrow}$-heavy-hole pair and a $\sigma_-$ photon creates a $\ket{\uparrow}$-electron and $\ket{\Downarrow}$-heavy-hole pair. However, in the general case, the eigenfunctions have mixed heavy-hole/light-hole character. In case of sufficient valence-band mixing, the $\ket{\Downarrow}$ heavy-hole state is a superposition involving some light-hole components which makes the previously forbidden transitions become optically weakly active \cite{kumarPRB06, belhadjAPL10}. The amount of mixing increases for decreasing splitting between the heavy-hole and light-hole bands, therefore the effective tensile strain we introduce could potentially increase the valence-band mixing. However, since the change in energy due to hole-burning is on the order of tens of $\mu$eV, while the splitting between the heavy-hole and light-hole valence bands is on the order of tens of meV, we do not expect the effect to be strong.

Second, strain has also effects on exchange interaction. In particular, the bright neutral exciton state consists of linear-polarized doublet, split by few tens of $\mu$eV, due to anisotropic electron-hole exchange interaction.  Recent theory and experiments showed that strain, as a symmetry-breaking effect, can affect the polarization direction of the excitonic emission and the magnitude of the fine-structure splitting \cite{bryantPRL10, plumhofArxiv10}.  Therefore, particular care should be used when employing strain-tuning techniques with neutral excitons. On the hand hand, the effect of strain on the trion singlet transition is expected to be negligible, since there is no anisotropic exchange interaction for a spin-zero singlet electron pair and a hole. Therefore, strain-tuning could be extremely powerful and safe for quantum information applications involving trions \cite{bonatoPRL2010}.

\section {Conclusions}
We have studied experimentally and theoretically how the strain introduced by laser-induced surface defects affects the optical transition of QDs embedded in micropillar cavities. We showed that the quantum dot transition red-shifts, independently of the actual position of the hole. The quantum dot frequency shift is found to be about five times larger than the shift of a micropillar cavity mode resonance. Each hole burnt results in a shift of about $50$-$100$ $\mu$eV. We discussed a simple theoretical model to explain the experimental data.\\
These results can be used for fine-tuning purposes \cite{gudatAPL11}. Briefly, the cavity mode can be reduced to polarization-degeneracy changing the local birefringence in the cavity area by hole-burning. The amount and direction of the birefringence strongly depends on the orientation of the hole with respect to the cavity. Once the mode is polarization-degenerate, a blue-detuned quantum dot transition can be brought into resonance with the cavity mode by burning pairs of holes along orthogonal directions. In this way, the birefringence in the cavity region is left unaffected, while the dot transition is red-shifted into resonance.\\
We believe this tool can be useful for the implementation of quantum information applications based on solid-state cavity-QED with self-assembled quantum dots.

\section*{Acknowledgments}
This work was supported by the NSF grant 0901886, the Marie-Curie award No. EXT-CT-2006-042580 and FOM$\backslash$NWO grant No. 09PR2721-2.


\begin{thebibliography}{56}
\expandafter\ifx\csname natexlab\endcsname\relax\def\natexlab#1{#1}\fi
\expandafter\ifx\csname bibnamefont\endcsname\relax
  \def\bibnamefont#1{#1}\fi
\expandafter\ifx\csname bibfnamefont\endcsname\relax
  \def\bibfnamefont#1{#1}\fi
\expandafter\ifx\csname citenamefont\endcsname\relax
  \def\citenamefont#1{#1}\fi
\expandafter\ifx\csname url\endcsname\relax
  \def\url#1{\texttt{#1}}\fi
\expandafter\ifx\csname urlprefix\endcsname\relax\def\urlprefix{URL }\fi
\providecommand{\bibinfo}[2]{#2}
\providecommand{\eprint}[2][]{\url{#2}}

\bibitem[{\citenamefont{Lounis and Orrit}(2005)}]{lounisRPP05}
\bibinfo{author}{\bibfnamefont{B.}~\bibnamefont{Lounis}} \bibnamefont{and}
  \bibinfo{author}{\bibfnamefont{M.}~\bibnamefont{Orrit}},
  \bibinfo{journal}{Rep. Progr. Phys.} \textbf{\bibinfo{volume}{68}},
  \bibinfo{pages}{1129} (\bibinfo{year}{2005}).

\bibitem[{\citenamefont{Scheel}(2009)}]{scheelJMO09}
\bibinfo{author}{\bibfnamefont{S.}~\bibnamefont{Scheel}}, \bibinfo{journal}{J.
  Mod. Opt.} \textbf{\bibinfo{volume}{56}}, \bibinfo{pages}{141}
  (\bibinfo{year}{2009}).

\bibitem[{\citenamefont{Strauf et~al.}(2007)\citenamefont{Strauf, Stoltz,
  Rakher, Coldren, Petroff, and Bouwmeester}}]{straufNP07}
\bibinfo{author}{\bibfnamefont{S.}~\bibnamefont{Strauf}},
  \bibinfo{author}{\bibfnamefont{N.~G.} \bibnamefont{Stoltz}},
  \bibinfo{author}{\bibfnamefont{M.~T.} \bibnamefont{Rakher}},
  \bibinfo{author}{\bibfnamefont{L.~A.} \bibnamefont{Coldren}},
  \bibinfo{author}{\bibfnamefont{P.~M.} \bibnamefont{Petroff}},
  \bibnamefont{and}
  \bibinfo{author}{\bibfnamefont{D.}~\bibnamefont{Bouwmeester}},
  \bibinfo{journal}{Nat. Photon.} \textbf{\bibinfo{volume}{1}},
  \bibinfo{pages}{704} (\bibinfo{year}{2007}).

\bibitem[{\citenamefont{Reitzenstein and Forchel}(2010)}]{forchelIOP10}
\bibinfo{author}{\bibfnamefont{S.}~\bibnamefont{Reitzenstein}}
  \bibnamefont{and} \bibinfo{author}{\bibfnamefont{A.}~\bibnamefont{Forchel}},
  \bibinfo{journal}{Journal of Physics D: Applied Physics}
  \textbf{\bibinfo{volume}{43}}, \bibinfo{pages}{033001}
  (\bibinfo{year}{2010}).

\bibitem[{\citenamefont{Heindel et~al.}(2010)\citenamefont{Heindel, Schneider,
  Lermer, Kwon, Braun, Reitzenstein, H\"{o}fling, Kamp, and
  Forchel}}]{heindelAPL10}
\bibinfo{author}{\bibfnamefont{T.}~\bibnamefont{Heindel}},
  \bibinfo{author}{\bibfnamefont{C.}~\bibnamefont{Schneider}},
  \bibinfo{author}{\bibfnamefont{M.}~\bibnamefont{Lermer}},
  \bibinfo{author}{\bibfnamefont{S.~H.} \bibnamefont{Kwon}},
  \bibinfo{author}{\bibfnamefont{T.}~\bibnamefont{Braun}},
  \bibinfo{author}{\bibfnamefont{S.}~\bibnamefont{Reitzenstein}},
  \bibinfo{author}{\bibfnamefont{S.}~\bibnamefont{H\"{o}fling}},
  \bibinfo{author}{\bibfnamefont{M.}~\bibnamefont{Kamp}}, \bibnamefont{and}
  \bibinfo{author}{\bibfnamefont{A.}~\bibnamefont{Forchel}},
  \bibinfo{journal}{Applied Physics Letters} \textbf{\bibinfo{volume}{96}},
  \bibinfo{eid}{011107} (\bibinfo{year}{2010}).

\bibitem[{\citenamefont{Imamoglu et~al.}(1999)}]{imamogluPRL99}
\bibinfo{author}{\bibfnamefont{A.}~\bibnamefont{Imamoglu}}
  \bibnamefont{et~al.}, \bibinfo{journal}{Phys. Rev. Lett.}
  \textbf{\bibinfo{volume}{83}}, \bibinfo{pages}{4202} (\bibinfo{year}{1999}).

\bibitem[{\citenamefont{Reithmaier et~al.}(2004)\citenamefont{Reithmaier, Sek,
  Loffler, Hofmann, Kuhn, Reitzenstein, Keldysh, Kulakovskii, Reinecke, and
  Forchel}}]{reithmayerNature04}
\bibinfo{author}{\bibfnamefont{J.~P.} \bibnamefont{Reithmaier}},
  \bibinfo{author}{\bibfnamefont{G.}~\bibnamefont{Sek}},
  \bibinfo{author}{\bibfnamefont{A.}~\bibnamefont{Loffler}},
  \bibinfo{author}{\bibfnamefont{C.}~\bibnamefont{Hofmann}},
  \bibinfo{author}{\bibfnamefont{S.}~\bibnamefont{Kuhn}},
  \bibinfo{author}{\bibfnamefont{S.}~\bibnamefont{Reitzenstein}},
  \bibinfo{author}{\bibfnamefont{L.~V.} \bibnamefont{Keldysh}},
  \bibinfo{author}{\bibfnamefont{V.~D.} \bibnamefont{Kulakovskii}},
  \bibinfo{author}{\bibfnamefont{T.~L.} \bibnamefont{Reinecke}},
  \bibnamefont{and} \bibinfo{author}{\bibfnamefont{A.}~\bibnamefont{Forchel}},
  \bibinfo{journal}{Nature} \textbf{\bibinfo{volume}{432}},
  \bibinfo{pages}{197} (\bibinfo{year}{2004}).

\bibitem[{\citenamefont{Rakher et~al.}(2009)\citenamefont{Rakher, Stoltz,
  Coldren, Petroff, and Bouwmeester}}]{rakherPRL09}
\bibinfo{author}{\bibfnamefont{M.~T.} \bibnamefont{Rakher}},
  \bibinfo{author}{\bibfnamefont{N.~G.} \bibnamefont{Stoltz}},
  \bibinfo{author}{\bibfnamefont{L.~A.} \bibnamefont{Coldren}},
  \bibinfo{author}{\bibfnamefont{P.~M.} \bibnamefont{Petroff}},
  \bibnamefont{and}
  \bibinfo{author}{\bibfnamefont{D.}~\bibnamefont{Bouwmeester}},
  \bibinfo{journal}{Phys. Rev. Lett.} \textbf{\bibinfo{volume}{102}},
  \bibinfo{pages}{097403} (\bibinfo{year}{2009}).

\bibitem[{\citenamefont{Hu et~al.}(2009)\citenamefont{Hu, Munro, O'Brien, and
  Rarity}}]{xuPRB09}
\bibinfo{author}{\bibfnamefont{C.~Y.} \bibnamefont{Hu}},
  \bibinfo{author}{\bibfnamefont{W.~J.} \bibnamefont{Munro}},
  \bibinfo{author}{\bibfnamefont{J.~L.} \bibnamefont{O'Brien}},
  \bibnamefont{and} \bibinfo{author}{\bibfnamefont{J.~G.}
  \bibnamefont{Rarity}}, \bibinfo{journal}{Phys. Rev. B}
  \textbf{\bibinfo{volume}{80}}, \bibinfo{pages}{205326}
  (\bibinfo{year}{2009}).

\bibitem[{\citenamefont{Bonato et~al.}(2010)\citenamefont{Bonato, Haupt,
  Oemrawsingh, Gudat, Ding, van Exter, and Bouwmeester}}]{bonatoPRL2010}
\bibinfo{author}{\bibfnamefont{C.}~\bibnamefont{Bonato}},
  \bibinfo{author}{\bibfnamefont{F.}~\bibnamefont{Haupt}},
  \bibinfo{author}{\bibfnamefont{S.~S.~R.} \bibnamefont{Oemrawsingh}},
  \bibinfo{author}{\bibfnamefont{J.}~\bibnamefont{Gudat}},
  \bibinfo{author}{\bibfnamefont{D.}~\bibnamefont{Ding}},
  \bibinfo{author}{\bibfnamefont{M.~P.} \bibnamefont{van Exter}},
  \bibnamefont{and}
  \bibinfo{author}{\bibfnamefont{D.}~\bibnamefont{Bouwmeester}},
  \bibinfo{journal}{Phys. Rev. Lett.} \textbf{\bibinfo{volume}{104}},
  \bibinfo{pages}{160503} (\bibinfo{year}{2010}).

\bibitem[{\citenamefont{Cirac et~al.}(1997)\citenamefont{Cirac, Zoller, Kimble,
  and Mabuchi}}]{ciracPRL97}
\bibinfo{author}{\bibfnamefont{J.~I.} \bibnamefont{Cirac}},
  \bibinfo{author}{\bibfnamefont{P.}~\bibnamefont{Zoller}},
  \bibinfo{author}{\bibfnamefont{H.~J.} \bibnamefont{Kimble}},
  \bibnamefont{and} \bibinfo{author}{\bibfnamefont{H.}~\bibnamefont{Mabuchi}},
  \bibinfo{journal}{Phys. Rev. Lett.} \textbf{\bibinfo{volume}{78}},
  \bibinfo{pages}{3221} (\bibinfo{year}{1997}).

\bibitem[{\citenamefont{van Enk et~al.}(1997)\citenamefont{van Enk, Cirac, and
  Zoller}}]{vanenkPRL97}
\bibinfo{author}{\bibfnamefont{S.~J.} \bibnamefont{van Enk}},
  \bibinfo{author}{\bibfnamefont{J.~I.} \bibnamefont{Cirac}}, \bibnamefont{and}
  \bibinfo{author}{\bibfnamefont{P.}~\bibnamefont{Zoller}},
  \bibinfo{journal}{Phys. Rev. Lett.} \textbf{\bibinfo{volume}{78}},
  \bibinfo{pages}{4293} (\bibinfo{year}{1997}).

\bibitem[{\citenamefont{Warburton}(2002)}]{warburtonCP02}
\bibinfo{author}{\bibfnamefont{R.~J.} \bibnamefont{Warburton}},
  \bibinfo{journal}{Contemp. Phys.} \textbf{\bibinfo{volume}{43}},
  \bibinfo{pages}{351} (\bibinfo{year}{2002}).

\bibitem[{\citenamefont{Michler}(2003)}]{michlerBook03}
\bibinfo{author}{\bibfnamefont{P.}~\bibnamefont{Michler}},
  \emph{\bibinfo{title}{Single Quantum dots. Fundamentals, applications, and
  new concepts.}} (\bibinfo{publisher}{Springer}, \bibinfo{year}{2003}).

\bibitem[{\citenamefont{Schneider et~al.}(2008)\citenamefont{Schneider,
  Strau\ss, S\"{u}nner, Huggenberger, Wiener, Reitzenstein, Kamp, H\"{o}fling,
  and Forchel}}]{schneiderAPL08}
\bibinfo{author}{\bibfnamefont{C.}~\bibnamefont{Schneider}},
  \bibinfo{author}{\bibfnamefont{M.}~\bibnamefont{Strau\ss}},
  \bibinfo{author}{\bibfnamefont{T.}~\bibnamefont{S\"{u}nner}},
  \bibinfo{author}{\bibfnamefont{A.}~\bibnamefont{Huggenberger}},
  \bibinfo{author}{\bibfnamefont{D.}~\bibnamefont{Wiener}},
  \bibinfo{author}{\bibfnamefont{S.}~\bibnamefont{Reitzenstein}},
  \bibinfo{author}{\bibfnamefont{M.}~\bibnamefont{Kamp}},
  \bibinfo{author}{\bibfnamefont{S.}~\bibnamefont{H\"{o}fling}},
  \bibnamefont{and} \bibinfo{author}{\bibfnamefont{A.}~\bibnamefont{Forchel}},
  \bibinfo{journal}{Applied Physics Letters} \textbf{\bibinfo{volume}{92}},
  \bibinfo{eid}{183101} (\bibinfo{year}{2008}).

\bibitem[{\citenamefont{Dousse et~al.}(2008)\citenamefont{Dousse, Lanco,
  Suffczy\ifmmode~\acute{n}\else \'{n}\fi{}ski, Semenova, Miard, Lema\^\i{}tre,
  Sagnes, Roblin, Bloch, and Senellart}}]{doussePRL08}
\bibinfo{author}{\bibfnamefont{A.}~\bibnamefont{Dousse}},
  \bibinfo{author}{\bibfnamefont{L.}~\bibnamefont{Lanco}},
  \bibinfo{author}{\bibfnamefont{J.}~\bibnamefont{Suffczy\ifmmode~\acute{n}\else
  \'{n}\fi{}ski}}, \bibinfo{author}{\bibfnamefont{E.}~\bibnamefont{Semenova}},
  \bibinfo{author}{\bibfnamefont{A.}~\bibnamefont{Miard}},
  \bibinfo{author}{\bibfnamefont{A.}~\bibnamefont{Lema\^\i{}tre}},
  \bibinfo{author}{\bibfnamefont{I.}~\bibnamefont{Sagnes}},
  \bibinfo{author}{\bibfnamefont{C.}~\bibnamefont{Roblin}},
  \bibinfo{author}{\bibfnamefont{J.}~\bibnamefont{Bloch}}, \bibnamefont{and}
  \bibinfo{author}{\bibfnamefont{P.}~\bibnamefont{Senellart}},
  \bibinfo{journal}{Phys. Rev. Lett.} \textbf{\bibinfo{volume}{101}},
  \bibinfo{pages}{267404} (\bibinfo{year}{2008}).

\bibitem[{\citenamefont{Thon et~al.}(2009)\citenamefont{Thon, Rakher, Kim,
  Gudat, Irvine, Petroff, and Bouwmeester}}]{thonAPL09}
\bibinfo{author}{\bibfnamefont{S.~M.} \bibnamefont{Thon}},
  \bibinfo{author}{\bibfnamefont{M.~T.} \bibnamefont{Rakher}},
  \bibinfo{author}{\bibfnamefont{H.}~\bibnamefont{Kim}},
  \bibinfo{author}{\bibfnamefont{J.}~\bibnamefont{Gudat}},
  \bibinfo{author}{\bibfnamefont{W.~T.~M.} \bibnamefont{Irvine}},
  \bibinfo{author}{\bibfnamefont{P.~M.} \bibnamefont{Petroff}},
  \bibnamefont{and}
  \bibinfo{author}{\bibfnamefont{D.}~\bibnamefont{Bouwmeester}},
  \bibinfo{journal}{Applied Physics Letters} \textbf{\bibinfo{volume}{94}},
  \bibinfo{eid}{111115} (\bibinfo{year}{2009}).

\bibitem[{\citenamefont{Badolato et~al.}()\citenamefont{Badolato, Hennessy,
  Atature, Dreiser, Hu, Petroff, and Imamoğlu}}]{badolatoScience05}
\bibinfo{author}{\bibfnamefont{A.}~\bibnamefont{Badolato}},
  \bibinfo{author}{\bibfnamefont{K.}~\bibnamefont{Hennessy}},
  \bibinfo{author}{\bibfnamefont{M.}~\bibnamefont{Atature}},
  \bibinfo{author}{\bibfnamefont{J.}~\bibnamefont{Dreiser}},
  \bibinfo{author}{\bibfnamefont{E.}~\bibnamefont{Hu}},
  \bibinfo{author}{\bibfnamefont{P.~M.} \bibnamefont{Petroff}},
  \bibnamefont{and}
  \bibinfo{author}{\bibfnamefont{A.}~\bibnamefont{Imamoğlu}},
  \bibinfo{journal}{Science}  (????).

\bibitem[{\citenamefont{Hennessy et~al.}(2005)\citenamefont{Hennessy, Badolato,
  Tamboli, Petroff, Hu, Atat\"{u}re, Dreiser, and
  Imamo\u{g}lu}}]{hennessyAPL05}
\bibinfo{author}{\bibfnamefont{K.}~\bibnamefont{Hennessy}},
  \bibinfo{author}{\bibfnamefont{A.}~\bibnamefont{Badolato}},
  \bibinfo{author}{\bibfnamefont{A.}~\bibnamefont{Tamboli}},
  \bibinfo{author}{\bibfnamefont{P.~M.} \bibnamefont{Petroff}},
  \bibinfo{author}{\bibfnamefont{E.}~\bibnamefont{Hu}},
  \bibinfo{author}{\bibfnamefont{M.}~\bibnamefont{Atat\"{u}re}},
  \bibinfo{author}{\bibfnamefont{J.}~\bibnamefont{Dreiser}}, \bibnamefont{and}
  \bibinfo{author}{\bibfnamefont{A.}~\bibnamefont{Imamo\u{g}lu}},
  \bibinfo{journal}{Applied Physics Letters} \textbf{\bibinfo{volume}{87}},
  \bibinfo{eid}{021108} (\bibinfo{year}{2005}).

\bibitem[{\citenamefont{Faraon et~al.}(2008)\citenamefont{Faraon, Englund,
  Bulla, Luther-Davies, Eggleton, Stoltz, Petroff, and
  Vu\v{c}kovi\'{c}}}]{faraonAPL08}
\bibinfo{author}{\bibfnamefont{A.}~\bibnamefont{Faraon}},
  \bibinfo{author}{\bibfnamefont{D.}~\bibnamefont{Englund}},
  \bibinfo{author}{\bibfnamefont{D.}~\bibnamefont{Bulla}},
  \bibinfo{author}{\bibfnamefont{B.}~\bibnamefont{Luther-Davies}},
  \bibinfo{author}{\bibfnamefont{B.~J.} \bibnamefont{Eggleton}},
  \bibinfo{author}{\bibfnamefont{N.}~\bibnamefont{Stoltz}},
  \bibinfo{author}{\bibfnamefont{P.}~\bibnamefont{Petroff}}, \bibnamefont{and}
  \bibinfo{author}{\bibfnamefont{J.}~\bibnamefont{Vu\v{c}kovi\'{c}}},
  \bibinfo{journal}{Applied Physics Letters} \textbf{\bibinfo{volume}{92}},
  \bibinfo{eid}{043123} (\bibinfo{year}{2008}).

\bibitem[{\citenamefont{Sridharan et~al.}(2010)\citenamefont{Sridharan, Waks,
  Solomon, and Fourkas}}]{sridharanAPL10}
\bibinfo{author}{\bibfnamefont{D.}~\bibnamefont{Sridharan}},
  \bibinfo{author}{\bibfnamefont{E.}~\bibnamefont{Waks}},
  \bibinfo{author}{\bibfnamefont{G.}~\bibnamefont{Solomon}}, \bibnamefont{and}
  \bibinfo{author}{\bibfnamefont{J.~T.} \bibnamefont{Fourkas}},
  \bibinfo{journal}{Applied Physics Letters} \textbf{\bibinfo{volume}{96}},
  \bibinfo{eid}{153303} (\bibinfo{year}{2010}).

\bibitem[{\citenamefont{Hennessy et~al.}(2006)\citenamefont{Hennessy,
  H\"{o}gerle, Hu, Badolato, and Imamo\u{g}lu}}]{hennessyAPL06}
\bibinfo{author}{\bibfnamefont{K.}~\bibnamefont{Hennessy}},
  \bibinfo{author}{\bibfnamefont{C.}~\bibnamefont{H\"{o}gerle}},
  \bibinfo{author}{\bibfnamefont{E.}~\bibnamefont{Hu}},
  \bibinfo{author}{\bibfnamefont{A.}~\bibnamefont{Badolato}}, \bibnamefont{and}
  \bibinfo{author}{\bibfnamefont{A.}~\bibnamefont{Imamo\u{g}lu}},
  \bibinfo{journal}{Applied Physics Letters} \textbf{\bibinfo{volume}{89}},
  \bibinfo{eid}{041118} (\bibinfo{year}{2006}).

\bibitem[{\citenamefont{Erickson et~al.}(2006)\citenamefont{Erickson, Rockwood,
  Emery, Scherer, and Psaltis}}]{ericksonOE06}
\bibinfo{author}{\bibfnamefont{D.}~\bibnamefont{Erickson}},
  \bibinfo{author}{\bibfnamefont{T.}~\bibnamefont{Rockwood}},
  \bibinfo{author}{\bibfnamefont{T.}~\bibnamefont{Emery}},
  \bibinfo{author}{\bibfnamefont{A.}~\bibnamefont{Scherer}}, \bibnamefont{and}
  \bibinfo{author}{\bibfnamefont{D.}~\bibnamefont{Psaltis}},
  \bibinfo{journal}{Opt. Lett.} \textbf{\bibinfo{volume}{31}},
  \bibinfo{pages}{59} (\bibinfo{year}{2006}).

\bibitem[{\citenamefont{Intonti et~al.}(2006)\citenamefont{Intonti, Vignolini,
  T\"{u}rck, Colocci, Bettotti, Pavesi, Schweizer, Wehrspohn, and
  Wiersma}}]{intontiAPL06}
\bibinfo{author}{\bibfnamefont{F.}~\bibnamefont{Intonti}},
  \bibinfo{author}{\bibfnamefont{S.}~\bibnamefont{Vignolini}},
  \bibinfo{author}{\bibfnamefont{V.}~\bibnamefont{T\"{u}rck}},
  \bibinfo{author}{\bibfnamefont{M.}~\bibnamefont{Colocci}},
  \bibinfo{author}{\bibfnamefont{P.}~\bibnamefont{Bettotti}},
  \bibinfo{author}{\bibfnamefont{L.}~\bibnamefont{Pavesi}},
  \bibinfo{author}{\bibfnamefont{S.~L.} \bibnamefont{Schweizer}},
  \bibinfo{author}{\bibfnamefont{R.}~\bibnamefont{Wehrspohn}},
  \bibnamefont{and} \bibinfo{author}{\bibfnamefont{D.}~\bibnamefont{Wiersma}},
  \bibinfo{journal}{Applied Physics Letters} \textbf{\bibinfo{volume}{89}},
  \bibinfo{eid}{211117} (\bibinfo{year}{2006}).

\bibitem[{\citenamefont{Mosor et~al.}(2005)\citenamefont{Mosor, Hendrickson,
  Richards, Sweet, Khitrova, Gibbs, Yoshie, Scherer, Shchekin, and
  Deppe}}]{mosorAPL05}
\bibinfo{author}{\bibfnamefont{S.}~\bibnamefont{Mosor}},
  \bibinfo{author}{\bibfnamefont{J.}~\bibnamefont{Hendrickson}},
  \bibinfo{author}{\bibfnamefont{B.~C.} \bibnamefont{Richards}},
  \bibinfo{author}{\bibfnamefont{J.}~\bibnamefont{Sweet}},
  \bibinfo{author}{\bibfnamefont{G.}~\bibnamefont{Khitrova}},
  \bibinfo{author}{\bibfnamefont{H.~M.} \bibnamefont{Gibbs}},
  \bibinfo{author}{\bibfnamefont{T.}~\bibnamefont{Yoshie}},
  \bibinfo{author}{\bibfnamefont{A.}~\bibnamefont{Scherer}},
  \bibinfo{author}{\bibfnamefont{O.~B.} \bibnamefont{Shchekin}},
  \bibnamefont{and} \bibinfo{author}{\bibfnamefont{D.~G.} \bibnamefont{Deppe}},
  \bibinfo{journal}{Applied Physics Letters} \textbf{\bibinfo{volume}{87}},
  \bibinfo{eid}{141105} (\bibinfo{year}{2005}).

\bibitem[{\citenamefont{Fry et~al.}(2000)\citenamefont{Fry, Itskevich, Mowbray,
  Skolnick, Finley, Barker, O'Reilly, Wilson, Larkin, Maksym
  et~al.}}]{fryPRL2000}
\bibinfo{author}{\bibfnamefont{P.~W.} \bibnamefont{Fry}},
  \bibinfo{author}{\bibfnamefont{I.~E.} \bibnamefont{Itskevich}},
  \bibinfo{author}{\bibfnamefont{D.~J.} \bibnamefont{Mowbray}},
  \bibinfo{author}{\bibfnamefont{M.~S.} \bibnamefont{Skolnick}},
  \bibinfo{author}{\bibfnamefont{J.~J.} \bibnamefont{Finley}},
  \bibinfo{author}{\bibfnamefont{J.~A.} \bibnamefont{Barker}},
  \bibinfo{author}{\bibfnamefont{E.~P.} \bibnamefont{O'Reilly}},
  \bibinfo{author}{\bibfnamefont{L.~R.} \bibnamefont{Wilson}},
  \bibinfo{author}{\bibfnamefont{I.~A.} \bibnamefont{Larkin}},
  \bibinfo{author}{\bibfnamefont{P.~A.} \bibnamefont{Maksym}},
  \bibnamefont{et~al.}, \bibinfo{journal}{Phys. Rev. Lett.}
  \textbf{\bibinfo{volume}{84}}, \bibinfo{pages}{733} (\bibinfo{year}{2000}).

\bibitem[{\citenamefont{Kistner et~al.}(2008)\citenamefont{Kistner, Heindel,
  Schneider, Rahimi-Iman, Reitzenstein, H\"{o}fling, and
  Forchel}}]{kistnerOE08}
\bibinfo{author}{\bibfnamefont{C.}~\bibnamefont{Kistner}},
  \bibinfo{author}{\bibfnamefont{T.}~\bibnamefont{Heindel}},
  \bibinfo{author}{\bibfnamefont{C.}~\bibnamefont{Schneider}},
  \bibinfo{author}{\bibfnamefont{A.}~\bibnamefont{Rahimi-Iman}},
  \bibinfo{author}{\bibfnamefont{S.}~\bibnamefont{Reitzenstein}},
  \bibinfo{author}{\bibfnamefont{S.}~\bibnamefont{H\"{o}fling}},
  \bibnamefont{and} \bibinfo{author}{\bibfnamefont{A.}~\bibnamefont{Forchel}},
  \bibinfo{journal}{Opt. Express} \textbf{\bibinfo{volume}{16}},
  \bibinfo{pages}{15006} (\bibinfo{year}{2008}).

\bibitem[{\citenamefont{Kiraz et~al.}(2001)\citenamefont{Kiraz, Michler,
  Becher, Gayral, Imamoglu, Zhang, Hu, Schoenfeld, and Petroff}}]{kirazAPL01}
\bibinfo{author}{\bibfnamefont{A.}~\bibnamefont{Kiraz}},
  \bibinfo{author}{\bibfnamefont{P.}~\bibnamefont{Michler}},
  \bibinfo{author}{\bibfnamefont{C.}~\bibnamefont{Becher}},
  \bibinfo{author}{\bibfnamefont{B.}~\bibnamefont{Gayral}},
  \bibinfo{author}{\bibfnamefont{A.}~\bibnamefont{Imamoglu}},
  \bibinfo{author}{\bibfnamefont{L.}~\bibnamefont{Zhang}},
  \bibinfo{author}{\bibfnamefont{E.}~\bibnamefont{Hu}},
  \bibinfo{author}{\bibfnamefont{W.~V.} \bibnamefont{Schoenfeld}},
  \bibnamefont{and} \bibinfo{author}{\bibfnamefont{P.~M.}
  \bibnamefont{Petroff}}, \bibinfo{journal}{Applied Physics Letters}
  \textbf{\bibinfo{volume}{78}}, \bibinfo{pages}{3932} (\bibinfo{year}{2001}).

\bibitem[{\citenamefont{Faraon et~al.}(2007)\citenamefont{Faraon, Englund,
  Fushman, Vu\v{c}kovi\'{c}, Stoltz, and Petroff}}]{faraonAPL07}
\bibinfo{author}{\bibfnamefont{A.}~\bibnamefont{Faraon}},
  \bibinfo{author}{\bibfnamefont{D.}~\bibnamefont{Englund}},
  \bibinfo{author}{\bibfnamefont{I.}~\bibnamefont{Fushman}},
  \bibinfo{author}{\bibfnamefont{J.}~\bibnamefont{Vu\v{c}kovi\'{c}}},
  \bibinfo{author}{\bibfnamefont{N.}~\bibnamefont{Stoltz}}, \bibnamefont{and}
  \bibinfo{author}{\bibfnamefont{P.}~\bibnamefont{Petroff}},
  \bibinfo{journal}{Applied Physics Letters} \textbf{\bibinfo{volume}{90}},
  \bibinfo{eid}{213110} (\bibinfo{year}{2007}).

\bibitem[{\citenamefont{Rastelli et~al.}(2007)\citenamefont{Rastelli, Ulhaq,
  Kiravittaya, Wang, Zrenner, and Schmidt}}]{rastelliAPL07}
\bibinfo{author}{\bibfnamefont{A.}~\bibnamefont{Rastelli}},
  \bibinfo{author}{\bibfnamefont{A.}~\bibnamefont{Ulhaq}},
  \bibinfo{author}{\bibfnamefont{S.}~\bibnamefont{Kiravittaya}},
  \bibinfo{author}{\bibfnamefont{L.}~\bibnamefont{Wang}},
  \bibinfo{author}{\bibfnamefont{A.}~\bibnamefont{Zrenner}}, \bibnamefont{and}
  \bibinfo{author}{\bibfnamefont{O.~G.} \bibnamefont{Schmidt}},
  \bibinfo{journal}{Applied Physics Letters} \textbf{\bibinfo{volume}{90}},
  \bibinfo{eid}{073120} (\bibinfo{year}{2007}).

\bibitem[{\citenamefont{Oberm\"{u}ller
  et~al.}(1999)\citenamefont{Oberm\"{u}ller, Deisenrieder, Abstreiter, Karrai,
  Grosse, Manus, Feldmann, Lipsanen, Sopanen, and Ahopelto}}]{obermullerAPL99}
\bibinfo{author}{\bibfnamefont{C.}~\bibnamefont{Oberm\"{u}ller}},
  \bibinfo{author}{\bibfnamefont{A.}~\bibnamefont{Deisenrieder}},
  \bibinfo{author}{\bibfnamefont{G.}~\bibnamefont{Abstreiter}},
  \bibinfo{author}{\bibfnamefont{K.}~\bibnamefont{Karrai}},
  \bibinfo{author}{\bibfnamefont{S.}~\bibnamefont{Grosse}},
  \bibinfo{author}{\bibfnamefont{S.}~\bibnamefont{Manus}},
  \bibinfo{author}{\bibfnamefont{J.}~\bibnamefont{Feldmann}},
  \bibinfo{author}{\bibfnamefont{H.}~\bibnamefont{Lipsanen}},
  \bibinfo{author}{\bibfnamefont{M.}~\bibnamefont{Sopanen}}, \bibnamefont{and}
  \bibinfo{author}{\bibfnamefont{J.}~\bibnamefont{Ahopelto}},
  \bibinfo{journal}{Applied Physics Letters} \textbf{\bibinfo{volume}{75}},
  \bibinfo{pages}{358} (\bibinfo{year}{1999}).

\bibitem[{\citenamefont{Zander et~al.}(2009)\citenamefont{Zander, Herklotz,
  Kiravittaya, Benyoucef, Ding, Atkinson, Kumar, Plumhof, D\"{o}rr, Rastelli
  et~al.}}]{zanderOE09}
\bibinfo{author}{\bibfnamefont{T.}~\bibnamefont{Zander}},
  \bibinfo{author}{\bibfnamefont{A.}~\bibnamefont{Herklotz}},
  \bibinfo{author}{\bibfnamefont{S.}~\bibnamefont{Kiravittaya}},
  \bibinfo{author}{\bibfnamefont{M.}~\bibnamefont{Benyoucef}},
  \bibinfo{author}{\bibfnamefont{F.}~\bibnamefont{Ding}},
  \bibinfo{author}{\bibfnamefont{P.}~\bibnamefont{Atkinson}},
  \bibinfo{author}{\bibfnamefont{S.}~\bibnamefont{Kumar}},
  \bibinfo{author}{\bibfnamefont{J.~D.} \bibnamefont{Plumhof}},
  \bibinfo{author}{\bibfnamefont{K.}~\bibnamefont{D\"{o}rr}},
  \bibinfo{author}{\bibfnamefont{A.}~\bibnamefont{Rastelli}},
  \bibnamefont{et~al.}, \bibinfo{journal}{Opt. Express}
  \textbf{\bibinfo{volume}{17}}, \bibinfo{pages}{22452} (\bibinfo{year}{2009}).

\bibitem[{\citenamefont{Bryant et~al.}(2010)\citenamefont{Bryant,
  Zieli\ifmmode~\acute{n}\else \'{n}\fi{}ski, Malkova, Sims, Jask\'olski, and
  Aizpurua}}]{bryantPRL10}
\bibinfo{author}{\bibfnamefont{G.~W.} \bibnamefont{Bryant}},
  \bibinfo{author}{\bibfnamefont{M.}~\bibnamefont{Zieli\ifmmode~\acute{n}\else
  \'{n}\fi{}ski}}, \bibinfo{author}{\bibfnamefont{N.}~\bibnamefont{Malkova}},
  \bibinfo{author}{\bibfnamefont{J.}~\bibnamefont{Sims}},
  \bibinfo{author}{\bibfnamefont{W.}~\bibnamefont{Jask\'olski}},
  \bibnamefont{and} \bibinfo{author}{\bibfnamefont{J.}~\bibnamefont{Aizpurua}},
  \bibinfo{journal}{Phys. Rev. Lett.} \textbf{\bibinfo{volume}{105}},
  \bibinfo{pages}{067404} (\bibinfo{year}{2010}).

\bibitem[{\citenamefont{Seidl et~al.}(2006)\citenamefont{Seidl, Kroner,
  H\"{o}gele, Karrai, Warburton, Badolato, and Petroff}}]{seidlAPL06}
\bibinfo{author}{\bibfnamefont{S.}~\bibnamefont{Seidl}},
  \bibinfo{author}{\bibfnamefont{M.}~\bibnamefont{Kroner}},
  \bibinfo{author}{\bibfnamefont{A.}~\bibnamefont{H\"{o}gele}},
  \bibinfo{author}{\bibfnamefont{K.}~\bibnamefont{Karrai}},
  \bibinfo{author}{\bibfnamefont{R.~J.} \bibnamefont{Warburton}},
  \bibinfo{author}{\bibfnamefont{A.}~\bibnamefont{Badolato}}, \bibnamefont{and}
  \bibinfo{author}{\bibfnamefont{P.~M.} \bibnamefont{Petroff}},
  \bibinfo{journal}{Applied Physics Letters} \textbf{\bibinfo{volume}{88}},
  \bibinfo{eid}{203113} (\bibinfo{year}{2006}).

\bibitem[{\citenamefont{van Doorn et~al.}(1996{\natexlab{a}})\citenamefont{van
  Doorn, van Exter, and Woerdman}}]{doornAPL96a}
\bibinfo{author}{\bibfnamefont{A.~K.~J.} \bibnamefont{van Doorn}},
  \bibinfo{author}{\bibfnamefont{M.~P.} \bibnamefont{van Exter}},
  \bibnamefont{and} \bibinfo{author}{\bibfnamefont{J.~P.}
  \bibnamefont{Woerdman}}, \bibinfo{journal}{Appl. Phys. Lett.}
  \textbf{\bibinfo{volume}{69}}, \bibinfo{pages}{1041}
  (\bibinfo{year}{1996}{\natexlab{a}}).

\bibitem[{\citenamefont{van Doorn et~al.}(1996{\natexlab{b}})\citenamefont{van
  Doorn, van Exter, and Woerdman}}]{doornAPL96b}
\bibinfo{author}{\bibfnamefont{A.~K.~J.} \bibnamefont{van Doorn}},
  \bibinfo{author}{\bibfnamefont{M.~P.} \bibnamefont{van Exter}},
  \bibnamefont{and} \bibinfo{author}{\bibfnamefont{J.~P.}
  \bibnamefont{Woerdman}}, \bibinfo{journal}{Appl. Phys. Lett.}
  \textbf{\bibinfo{volume}{69}}, \bibinfo{pages}{3635}
  (\bibinfo{year}{1996}{\natexlab{b}}).

\bibitem[{\citenamefont{Bonato et~al.}(2009)\citenamefont{Bonato, Ding, Gudat,
  Thon, Kim, Petroff, van Exter, and Bouwmeester}}]{bonatoAPL2009}
\bibinfo{author}{\bibfnamefont{C.}~\bibnamefont{Bonato}},
  \bibinfo{author}{\bibfnamefont{D.}~\bibnamefont{Ding}},
  \bibinfo{author}{\bibfnamefont{J.}~\bibnamefont{Gudat}},
  \bibinfo{author}{\bibfnamefont{S.}~\bibnamefont{Thon}},
  \bibinfo{author}{\bibfnamefont{H.}~\bibnamefont{Kim}},
  \bibinfo{author}{\bibfnamefont{P.~M.} \bibnamefont{Petroff}},
  \bibinfo{author}{\bibfnamefont{M.~P.} \bibnamefont{van Exter}},
  \bibnamefont{and}
  \bibinfo{author}{\bibfnamefont{D.}~\bibnamefont{Bouwmeester}},
  \bibinfo{journal}{Applied Physics Letters} \textbf{\bibinfo{volume}{95}},
  \bibinfo{eid}{251104} (\bibinfo{year}{2009}).

\bibitem[{\citenamefont{Gudat et~al.}(2011)\citenamefont{Gudat, Bonato, van
  Nieuwenburg, Thon, Kim, Petroff, van Exter, and Bouwmeester}}]{gudatAPL11}
\bibinfo{author}{\bibfnamefont{J.}~\bibnamefont{Gudat}},
  \bibinfo{author}{\bibfnamefont{C.}~\bibnamefont{Bonato}},
  \bibinfo{author}{\bibfnamefont{E.}~\bibnamefont{van Nieuwenburg}},
  \bibinfo{author}{\bibfnamefont{S.}~\bibnamefont{Thon}},
  \bibinfo{author}{\bibfnamefont{H.}~\bibnamefont{Kim}},
  \bibinfo{author}{\bibfnamefont{P.~M.} \bibnamefont{Petroff}},
  \bibinfo{author}{\bibfnamefont{M.~P.} \bibnamefont{van Exter}},
  \bibnamefont{and}
  \bibinfo{author}{\bibfnamefont{D.}~\bibnamefont{Bouwmeester}},
  \bibinfo{journal}{Applied Physics Letters} \textbf{\bibinfo{volume}{98}},
  \bibinfo{eid}{121111} (\bibinfo{year}{2011}).

\bibitem[{\citenamefont{Gudat et~al.}()\citenamefont{Gudat, Bonato, de~Vries,
  Thon, Kim, Petroff, van Exter, and Bouwmeester}}]{gudatNewAPL}
\bibinfo{author}{\bibfnamefont{J.}~\bibnamefont{Gudat}},
  \bibinfo{author}{\bibfnamefont{C.}~\bibnamefont{Bonato}},
  \bibinfo{author}{\bibfnamefont{K.}~\bibnamefont{de~Vries}},
  \bibinfo{author}{\bibfnamefont{S.}~\bibnamefont{Thon}},
  \bibinfo{author}{\bibfnamefont{H.}~\bibnamefont{Kim}},
  \bibinfo{author}{\bibfnamefont{P.~M.} \bibnamefont{Petroff}},
  \bibinfo{author}{\bibfnamefont{M.~P.} \bibnamefont{van Exter}},
  \bibnamefont{and}
  \bibinfo{author}{\bibfnamefont{D.}~\bibnamefont{Bouwmeester}},
  \bibinfo{note}{submitted}.

\bibitem[{\citenamefont{Stoltz et~al.}(2005)\citenamefont{Stoltz, Rakher,
  Strauf, Badolato, Lofgreen, Petroff, Coldren, and Bouwmeester}}]{stoltzAPL05}
\bibinfo{author}{\bibfnamefont{N.~G.} \bibnamefont{Stoltz}},
  \bibinfo{author}{\bibfnamefont{M.}~\bibnamefont{Rakher}},
  \bibinfo{author}{\bibfnamefont{S.}~\bibnamefont{Strauf}},
  \bibinfo{author}{\bibfnamefont{A.}~\bibnamefont{Badolato}},
  \bibinfo{author}{\bibfnamefont{D.~D.} \bibnamefont{Lofgreen}},
  \bibinfo{author}{\bibfnamefont{P.~M.} \bibnamefont{Petroff}},
  \bibinfo{author}{\bibfnamefont{L.~A.} \bibnamefont{Coldren}},
  \bibnamefont{and}
  \bibinfo{author}{\bibfnamefont{D.}~\bibnamefont{Bouwmeester}},
  \bibinfo{journal}{Appl. Phys. Lett.} \textbf{\bibinfo{volume}{87}},
  \bibinfo{eid}{031105} (\bibinfo{year}{2005}).

\bibitem[{\citenamefont{Warburton et~al.}(2000)\citenamefont{Warburton,
  Schaeflein, Haft, Bickel, Lorke, Karrai, Garcia, Schoenfeld, and
  Petroff}}]{warburtonNat00}
\bibinfo{author}{\bibfnamefont{R.~J.} \bibnamefont{Warburton}},
  \bibinfo{author}{\bibfnamefont{C.}~\bibnamefont{Schaeflein}},
  \bibinfo{author}{\bibfnamefont{D.}~\bibnamefont{Haft}},
  \bibinfo{author}{\bibfnamefont{F.}~\bibnamefont{Bickel}},
  \bibinfo{author}{\bibfnamefont{A.}~\bibnamefont{Lorke}},
  \bibinfo{author}{\bibfnamefont{K.}~\bibnamefont{Karrai}},
  \bibinfo{author}{\bibfnamefont{J.~M.} \bibnamefont{Garcia}},
  \bibinfo{author}{\bibfnamefont{M.}~\bibnamefont{Schoenfeld}},
  \bibnamefont{and} \bibinfo{author}{\bibfnamefont{P.~M.}
  \bibnamefont{Petroff}}, \bibinfo{journal}{Nature}
  \textbf{\bibinfo{volume}{405}}, \bibinfo{pages}{926} (\bibinfo{year}{2000}).

\bibitem[{\citenamefont{van Doorn et~al.}(1998)\citenamefont{van Doorn, van
  Exter, and Woerdman}}]{doornIEEE98}
\bibinfo{author}{\bibfnamefont{A.~K.~J.} \bibnamefont{van Doorn}},
  \bibinfo{author}{\bibfnamefont{M.~P.} \bibnamefont{van Exter}},
  \bibnamefont{and} \bibinfo{author}{\bibfnamefont{J.~P.}
  \bibnamefont{Woerdman}}, \bibinfo{journal}{IEEE Journ. of Quant. Electr.}
  \textbf{\bibinfo{volume}{34}}, \bibinfo{pages}{760} (\bibinfo{year}{1998}).

\bibitem[{\citenamefont{Chuang}(1995)}]{chuangBook}
\bibinfo{author}{\bibfnamefont{S.~L.} \bibnamefont{Chuang}},
  \emph{\bibinfo{title}{Physics of Optoelectronic Devices}}
  (\bibinfo{publisher}{Wiley}, \bibinfo{year}{1995}).

\bibitem[{\citenamefont{Watson et~al.}(2004)\citenamefont{Watson, Poirier,
  Heaton, Lewis, and Boudreau}}]{watsonJLT04}
\bibinfo{author}{\bibfnamefont{C.~D.} \bibnamefont{Watson}},
  \bibinfo{author}{\bibfnamefont{M.}~\bibnamefont{Poirier}},
  \bibinfo{author}{\bibfnamefont{J.~M.} \bibnamefont{Heaton}},
  \bibinfo{author}{\bibfnamefont{M.}~\bibnamefont{Lewis}}, \bibnamefont{and}
  \bibinfo{author}{\bibfnamefont{M.}~\bibnamefont{Boudreau}},
  \bibinfo{journal}{Journ. of Lightw. Tech.} \textbf{\bibinfo{volume}{22}},
  \bibinfo{pages}{1598} (\bibinfo{year}{2004}).

\bibitem[{\citenamefont{Andreani et~al.}(1987)\citenamefont{Andreani,
  Pasquarello, and Bassani}}]{andreaniPRB87}
\bibinfo{author}{\bibfnamefont{L.~C.} \bibnamefont{Andreani}},
  \bibinfo{author}{\bibfnamefont{A.}~\bibnamefont{Pasquarello}},
  \bibnamefont{and} \bibinfo{author}{\bibfnamefont{F.}~\bibnamefont{Bassani}},
  \bibinfo{journal}{Phys. Rev. B} \textbf{\bibinfo{volume}{36}},
  \bibinfo{pages}{5887} (\bibinfo{year}{1987}).

\bibitem[{\citenamefont{Grundmann et~al.}(1995)\citenamefont{Grundmann, Stier,
  and Bimberg}}]{grundmannPRB95}
\bibinfo{author}{\bibfnamefont{M.}~\bibnamefont{Grundmann}},
  \bibinfo{author}{\bibfnamefont{O.}~\bibnamefont{Stier}}, \bibnamefont{and}
  \bibinfo{author}{\bibfnamefont{D.}~\bibnamefont{Bimberg}},
  \bibinfo{journal}{Phys. Rev. B} \textbf{\bibinfo{volume}{52}},
  \bibinfo{pages}{11969} (\bibinfo{year}{1995}).

\bibitem[{\citenamefont{Li et~al.}(1996)\citenamefont{Li, Xia, Yuan, Xu, Ge,
  Wang, Wang, Wang, and Chang}}]{liPRB96}
\bibinfo{author}{\bibfnamefont{S.-S.} \bibnamefont{Li}},
  \bibinfo{author}{\bibfnamefont{J.-B.} \bibnamefont{Xia}},
  \bibinfo{author}{\bibfnamefont{Z.~L.} \bibnamefont{Yuan}},
  \bibinfo{author}{\bibfnamefont{Z.~Y.} \bibnamefont{Xu}},
  \bibinfo{author}{\bibfnamefont{W.}~\bibnamefont{Ge}},
  \bibinfo{author}{\bibfnamefont{X.~R.} \bibnamefont{Wang}},
  \bibinfo{author}{\bibfnamefont{Y.}~\bibnamefont{Wang}},
  \bibinfo{author}{\bibfnamefont{J.}~\bibnamefont{Wang}}, \bibnamefont{and}
  \bibinfo{author}{\bibfnamefont{L.~L.} \bibnamefont{Chang}},
  \bibinfo{journal}{Phys. Rev. B} \textbf{\bibinfo{volume}{54}},
  \bibinfo{pages}{11575} (\bibinfo{year}{1996}).

\bibitem[{\citenamefont{Tadi\ifmmode~\acute{c}\else \'{c}\fi{}
  et~al.}(2002)\citenamefont{Tadi\ifmmode~\acute{c}\else \'{c}\fi{}, Peeters,
  and Janssens}}]{tadicPRB02}
\bibinfo{author}{\bibfnamefont{M.}~\bibnamefont{Tadi\ifmmode~\acute{c}\else
  \'{c}\fi{}}}, \bibinfo{author}{\bibfnamefont{F.~M.} \bibnamefont{Peeters}},
  \bibnamefont{and} \bibinfo{author}{\bibfnamefont{K.~L.}
  \bibnamefont{Janssens}}, \bibinfo{journal}{Phys. Rev. B}
  \textbf{\bibinfo{volume}{65}}, \bibinfo{pages}{165333}
  (\bibinfo{year}{2002}).

\bibitem[{\citenamefont{Bryant}(1988)}]{bryantPRB1988}
\bibinfo{author}{\bibfnamefont{G.~W.} \bibnamefont{Bryant}},
  \bibinfo{journal}{Phys. Rev. B} \textbf{\bibinfo{volume}{37}},
  \bibinfo{pages}{8763} (\bibinfo{year}{1988}).

\bibitem[{\citenamefont{Franceschetti and Zunger}(1997)}]{franceschettiPRL1997}
\bibinfo{author}{\bibfnamefont{A.}~\bibnamefont{Franceschetti}}
  \bibnamefont{and} \bibinfo{author}{\bibfnamefont{A.}~\bibnamefont{Zunger}},
  \bibinfo{journal}{Phys. Rev. Lett.} \textbf{\bibinfo{volume}{78}},
  \bibinfo{pages}{915} (\bibinfo{year}{1997}).

\bibitem[{\citenamefont{Nazir et~al.}(2005)\citenamefont{Nazir, Lovett,
  Barrett, Reina, and Briggs}}]{nazirPRB2005}
\bibinfo{author}{\bibfnamefont{A.}~\bibnamefont{Nazir}},
  \bibinfo{author}{\bibfnamefont{B.~W.} \bibnamefont{Lovett}},
  \bibinfo{author}{\bibfnamefont{S.~D.} \bibnamefont{Barrett}},
  \bibinfo{author}{\bibfnamefont{J.~H.} \bibnamefont{Reina}}, \bibnamefont{and}
  \bibinfo{author}{\bibfnamefont{G.~A.~D.} \bibnamefont{Briggs}},
  \bibinfo{journal}{Phys. Rev. B} \textbf{\bibinfo{volume}{71}},
  \bibinfo{pages}{045334} (\bibinfo{year}{2005}).

\bibitem[{\citenamefont{Biolatti et~al.}(2002)\citenamefont{Biolatti, D'Amico,
  Zanardi, and Rossi}}]{biolattiPRB02}
\bibinfo{author}{\bibfnamefont{E.}~\bibnamefont{Biolatti}},
  \bibinfo{author}{\bibfnamefont{I.}~\bibnamefont{D'Amico}},
  \bibinfo{author}{\bibfnamefont{P.}~\bibnamefont{Zanardi}}, \bibnamefont{and}
  \bibinfo{author}{\bibfnamefont{F.}~\bibnamefont{Rossi}},
  \bibinfo{journal}{Phys. Rev. B} \textbf{\bibinfo{volume}{65}},
  \bibinfo{pages}{075306} (\bibinfo{year}{2002}).

\bibitem[{\citenamefont{Riel}(2008)}]{rielAJP08}
\bibinfo{author}{\bibfnamefont{B.~J.} \bibnamefont{Riel}},
  \bibinfo{journal}{American Journal of Physics} \textbf{\bibinfo{volume}{76}},
  \bibinfo{pages}{750} (\bibinfo{year}{2008}).

\bibitem[{\citenamefont{Kumar et~al.}(2006)\citenamefont{Kumar, Kapoor, Gupta,
  and Sen}}]{kumarPRB06}
\bibinfo{author}{\bibfnamefont{J.}~\bibnamefont{Kumar}},
  \bibinfo{author}{\bibfnamefont{S.}~\bibnamefont{Kapoor}},
  \bibinfo{author}{\bibfnamefont{S.~K.} \bibnamefont{Gupta}}, \bibnamefont{and}
  \bibinfo{author}{\bibfnamefont{P.~K.} \bibnamefont{Sen}},
  \bibinfo{journal}{Phys. Rev. B} \textbf{\bibinfo{volume}{74}},
  \bibinfo{pages}{115326} (\bibinfo{year}{2006}).

\bibitem[{\citenamefont{Belhadj et~al.}(2010)\citenamefont{Belhadj, Amand,
  Kunold, Simon, Kuroda, Abbarchi, Mano, Sakoda, Kunz, Marie
  et~al.}}]{belhadjAPL10}
\bibinfo{author}{\bibfnamefont{T.}~\bibnamefont{Belhadj}},
  \bibinfo{author}{\bibfnamefont{T.}~\bibnamefont{Amand}},
  \bibinfo{author}{\bibfnamefont{A.}~\bibnamefont{Kunold}},
  \bibinfo{author}{\bibfnamefont{C.-M.} \bibnamefont{Simon}},
  \bibinfo{author}{\bibfnamefont{T.}~\bibnamefont{Kuroda}},
  \bibinfo{author}{\bibfnamefont{M.}~\bibnamefont{Abbarchi}},
  \bibinfo{author}{\bibfnamefont{T.}~\bibnamefont{Mano}},
  \bibinfo{author}{\bibfnamefont{K.}~\bibnamefont{Sakoda}},
  \bibinfo{author}{\bibfnamefont{S.}~\bibnamefont{Kunz}},
  \bibinfo{author}{\bibfnamefont{X.}~\bibnamefont{Marie}},
  \bibnamefont{et~al.}, \bibinfo{journal}{Applied Physics Letters}
  \textbf{\bibinfo{volume}{97}}, \bibinfo{eid}{051111} (\bibinfo{year}{2010}).

\bibitem[{\citenamefont{Plumhof et~al.}()\citenamefont{Plumhof, Krapek, Ding,
  Joens, Hafenbrak, Klenovsky, Herklotz, Doerr, Michler, Rastelli
  et~al.}}]{plumhofArxiv10}
\bibinfo{author}{\bibfnamefont{J.~D.} \bibnamefont{Plumhof}},
  \bibinfo{author}{\bibfnamefont{V.}~\bibnamefont{Krapek}},
  \bibinfo{author}{\bibfnamefont{F.}~\bibnamefont{Ding}},
  \bibinfo{author}{\bibfnamefont{K.~D.} \bibnamefont{Joens}},
  \bibinfo{author}{\bibfnamefont{R.}~\bibnamefont{Hafenbrak}},
  \bibinfo{author}{\bibfnamefont{P.}~\bibnamefont{Klenovsky}},
  \bibinfo{author}{\bibfnamefont{A.}~\bibnamefont{Herklotz}},
  \bibinfo{author}{\bibfnamefont{K.}~\bibnamefont{Doerr}},
  \bibinfo{author}{\bibfnamefont{P.}~\bibnamefont{Michler}},
  \bibinfo{author}{\bibfnamefont{A.}~\bibnamefont{Rastelli}},
  \bibnamefont{et~al.}, \bibinfo{note}{arxiv:1011.3003}.

\end{thebibliography}

\end{document}